\documentclass[12pt]{elsarticle}
\usepackage{amssymb}
\usepackage{amsmath}
\usepackage{bm}
\usepackage{subfigure}
\usepackage{float}
\usepackage{natbib}
\usepackage{lscape}
\usepackage{multirow}
\usepackage{array,multirow}
\biboptions{sort&compress}
\begin{document}
\begin{frontmatter}
\title{Densities mixture unfolding for data obtained from detectors
  with finite resolution and limited acceptance}
\author{N.D. Gagunashvili\corref{cor1}\fnref{fn1}}
\fntext[fn1]{Present address: Max-Planck-Institut f\"{u}r Kernphysik, PO Box
103980, \\ 69029 Heidelberg, Germany}
\ead{nikolai@unak.is}
\cortext[cor1]{Tel.: +354-4608505; fax: +354-4608998}
\address{University of Akureyri, Borgir, v/Nordursl\'od, IS-600 Akureyri, Iceland}
\begin{abstract}
  A procedure based on a Mixture Density Model for correcting experimental
  data for distortions due to finite resolution and limited detector
  acceptance is presented. Addressing the case that the solution is known
  to be non-negative, in the approach presented here, the true distribution
  is estimated by a weighted sum of probability density functions with
  positive weights and with the width of the densities acting as a
  regularisation parameter responsible for the smoothness of the result.
  To obtain better smoothing in less populated regions, the width parameter
   is chosen inversely proportional to the square root of the
   estimated density. Furthermore,
  the non-negative garrote method is used to find the most economic
  representation of the solution. Cross-validation is employed to
  determine the optimal values of the resolution and garrote parameters.
  The proposed approach is directly applicable to multidimensional
  problems. Numerical examples in one and two dimensions are presented to
  illustrate the procedure.
\end{abstract}
\end{frontmatter}

\section{Introduction}
The probability density function (PDF) $P(x')$ of an experimentally
measured characteristic $x'$,  in general,  differs from the true physical
PDF $p(x)$\/ because of the limited acceptance (probability) $A(x)$  to
register an event with true characteristic $x$, finite resolution and
bias in the response function $R(x'|x)$, which describes the probability
to observe $x'$\/ for a given true value $x$. Formally the relation
between $P(x')$\/ and $p(x)$\/ is given by
\begin{equation}
    P(x') \propto \int_{\Omega} p(x)A(x)R(x'|x) \,dx \;.
\label{p1_main}
\end{equation}
The integration in (\ref{p1_main}) is carried out over the domain
$\Omega$\/ of the variable $x$. In practical applications the experimental
distribution is usually discretised by using a histogram representation,
obtained by integrating $P(x')$\/ over $n$\/ finite sized bins
\begin{equation}
    P_{j} = \int_{c_{j-1}}^{c_j} P(x') dx' \quad j=1,\ldots,n
\end{equation}
with $c_{j-1}, c_j$\/ the limits of bin $j$.

If a parametric (theoretical) model $p(x,a_{1},a_{2},\ldots, a_{l})$\/ for the true
PDF is known, then the unfolding can be done by determining the parameters. For
example, by a least squares fit to the binned data \cite{fitgagunash,zechebook,zhigunov}.
Here the model, which allows to describe the true distribution by a finite
number of parameter values, constitutes a priori information which is
needed to correct for the distortions by the experimental setup,

In contrast, model independent unfolding, as considered e.g. in \cite{zhigunov2,
blobel,correcting,gagunashvili,schmelling,hocker,zech,agost,blobel2,
gagunashvili_phystat,albert}, is an ill-posed problem, and every approach
to solve it requires a priori information about the solution. Methods differ,
directly or indirectly, in the way a priori information is incorporated in
the result.

\section{Description of the unfolding method}
\label{sec:method}
To solve the unfolding problem (\ref{p1_main}),  a representation of the
true distribution has to be chosen. This representation should be as
flexible as possible and  allow introducing a priori information.
Classical kernel statistics is an example that approximates the
true distribution by putting a $1/N$-weighed copy of a kernel PDF
at the location of each of $N$\/ observed data points and adding them up
(see e.g. \cite{silverman}). With enough data, this comes arbitrarily
close to any PDF. There exist methods that
use a kernel representation of the true distribution to solve also the
inverse problem \cite{kernel}. One drawback of this approach is that
one has to store all the data points, another is that the known
kernel based algorithms expect the response function of a set-up in
analytical form, i.e. computer modelling cannot be used.

In this paper the use of a Mixture Density Model (MDM) \cite{mixture,shalizi}
to describe the true distribution $p(x)$\/ is proposed,
\begin{equation}
\label{repres}
  p(x)=  \sum_{i=1}^{s} w_i \, K_i(x; a_{1i},...,a_{li}),
\end{equation}
where the $K_i(x; a_{1i},...,a_{li})$ is the $i$th Probability Density Function in Mixture (PDFM)
with parameters $a_{1i},...,a_{li}$\/ and the weight $w_i$\/ the fraction
of the $i$th PDFM.

The MDM lies between the cases of the parametric  representation of
the true density on one hand, i.e. the case when there is only one
distribution in the sum (\ref{repres}), and the kernel statistics
approach where the number of terms in the sum (\ref{repres}) is
equal to the number of observations $N$. The MDM has a limited
number of parameters for representing a PDF and computer modelling
can be used to calculate the response of the system. The MDM is also
 convenient for taking into account different type of a priori
information, such as knowledge about the type of distributions,
constraints on parameters, smoothness of the distributions and so on.
Ideas and achievements of regression analysis as well as classical
kernel statistics can be used in applications of a MDM for
estimating the densities.

Using Eq.\,(\ref{repres}) to parameterise the solution $p(x)$\/
reduces the unfolding problem from finding a solution in the
infinite-dimensional space of all functions to finding a solution
in a finite dimensional space. This way an approximation of the true
density is performed which, in contrast to e.g. a discretisation by
a histogram, has the advantage to introduce negligible quantisation
errors for sufficiently smooth distributions.

Without loss of generality two-parametric PDFMs will be used throughout the paper. The first parameter, $x_i$, defines the mean value (location) of term $i$\/ and the second one, $\lambda_i$, represents the standard deviation. Different smooth PDFMs commonly employed by kernel statistics, such as biweight, triweight, tricube, cosine, Cauchy, B-spline and other kernels can be
used. Rather popular is the Gaussian Mixture Model (GMM) \cite{gmm}  with PDFMs
\begin{equation}
   K_i(x; x_{i},\lambda_{i})
 = \frac{1}{\lambda_{i} \sqrt{2 \pi}}
   \exp\left({-\frac{(x-x_{i})^2}{2\lambda_{i}^2}}\right) \;,
\end{equation}
which provides a rather flexible model in the approximation of a wide
class of statistical distributions. The standard deviation $\lambda_{i}$\/
acts as a regularisation parameter, which allows to adjust the smoothness
of the result. Weights, positions $x_{i}$\/ and standard deviations
$\lambda_{i}$\/ are determined by the unfolding procedure described below.

Substituting $p(x)$\/ as represented by Eq\,(\ref{repres}) into
the basic Eq.\,(\ref{p1_main}) yields
\begin{equation}
 P(x') = \sum_{i=1}^{s} w_i \;
         \int_{\Omega} K_i(x;x_{i},\lambda_{i}) A(x)R(x'|x) \,dx \;,
\end{equation}
and taking statistical fluctuations into account, the relation between
the weights $w_i$\/ and the histogram of the observed distribution
becomes a set of linear equations
\begin{equation}
\label{basicp}
 \bm{P} =  \bm{\mathrm{Q}{w}}+  \bm{\epsilon} \;,
\end{equation}
where $\bm{P}$\/ is the $n$-component column vector of the experimentally
measured histogram, $\bm{w} = (w_1,w_2,...,w_s)^t$\/ is the $s$-component
vector of weights and $\bm{\mathrm{Q}}$\/ is an $n \times s$\/ matrix
with elements
\begin{equation}
   Q_{ji} = \int_{c_{j-1}}^{c_j} K_i(x; x_i, \lambda_i )\,A(x)\,R(x'|x)dx
   \quad j=1,\ldots, n; \quad i=1,\ldots, s \;.
\end{equation}
The vector $\bm{\epsilon}$\/ is an $n$-component vector of random deviates
with expectation value $E[\bm{\epsilon}]=\bm{0}$\/ and covariance matrix
$\bm{\mathrm{C}}$, the diagonal elements of which being
$\mathrm{Var}[\bm{\epsilon}]=\mathrm{diag}(\sigma_1^2,\sigma_2^2,\cdots,\sigma_n^2)$,
where $\sigma_j$\/ is the statistical error of the measured distribution
for the $j$th bin. Each column of the matrix $\bm{\mathrm{Q}}$\/ is the
response of the system to one of the PDFM in the mixture model for the
true distribution. Numerically the calculation of the column vectors can be
done by weighting events of a Monte Carlo sample such that they follow the
corresponding PDFM, see Ref.~\cite{sobol}, and taking the histogram of the
observed distribution obtained with the weighted entries.

By a non-negative least-squares fit, the weight vector $\bm{w}$\/ in
Eq.\,(\ref{basicp}) for a given set of PDFMs is determined such that it minimizes
\begin{equation}
\label{xieq}
   X^2 = (\bm{P} -\bm{\mathrm{Q}} \hat{\bm{w}})^t
         \bm{\mathrm{C}}^{-1}
        (\bm{P} - \bm{\mathrm{Q}}\hat{\bm{w}})
\end{equation}
under the constraints
\begin{equation}
\label{ineq}
w_i \geq 0 \quad i=1,...,s \;.
\end{equation}
Following reference \cite{lawson}, if an unconstrained solution  satisfies
Eq.\,(\ref{ineq}) then $\hat{\bm{w}}$\/ solves the constrained  problem.
Otherwise, the solution to the constrained problem must be a boundary point
of $[0,+\infty)^s$\/ and therefore at least one $w_i=0$. It follows that
after performing all possible regressions with one or more $w_i$\/ in
Eq.\,(\ref{ineq}) set to zero, the non-negative problem is solved by picking
the subset of $w_i$\/ satisfying Eq.\,(\ref{ineq}) such that $X^2$\/ as defined
in Eq.(\ref{xieq}) is smallest. The numerical algorithm and computer program for solving this minimisation problem has been developed in references \cite{lawson,lawsonpr}. Here, first the subset of components equal to zero is determined iteratively, and the vector of the remaining indices $\hat{\bm{w}}$\/ is found by simple linear regression
\begin{equation}
\label{eq:lsq}
  \hat{\bm{w}}
     =  (\bm{\mathcal Q^t \bm{\mathrm{C}}^{-1} \mathcal Q})^{-1}\,
        (\bm{\mathcal Q^t \bm{\mathrm{C}}^{-1}})\, \bm{P} \;,
\end{equation}
where $\bm{\mathcal Q}$\/ is the submatrix of $\bm{\mathrm{Q}}$\/ that
corresponds to the subset of indices of positive components of the solution.
The result of the fit is an estimate of the unfolded distribution $\hat{p}(x)$,
defined by a subset of parameters $x_{i}, \lambda_i$, $i=1,\ldots,k$\/ which
are summed with positive weights $\hat w_i,i=1,\ldots,k$\/ to yield
\begin{equation}
  \hat p(x)= \sum_{i=1}^{k} \hat{w}_i K_i(x; x_i,\lambda_i) \label{px}  \;.
\end{equation}

The choices of the optimal type of PDFMs and the values of parameters
(mean values and the standard deviations for the GMM model) are driven by the
accuracy and the complexity of the model. The goal is a simple, and at the same
time, accurate solution of the problem. A figure of merit for the accuracy is the
Prediction Error ($PE$) \cite{breiman}, defined as the expectation value of
the average squared normalised residual when using the predictor
$\bm{\mathcal{Q}{\hat w}}$\/ to describe an independent experimentally
measured histogram $\bm{P}^{new}$\/ drawn from the same parent distribution
as the original,
\begin{equation}
   PE(\bm{\mathcal{Q}{\hat w}})
 = E[\frac{1}{n}(\bm{P}^{new}-\bm{\mathcal{Q}}\hat{\bm{w}})^t
                 \bm{\mathrm{C}}^{-1}
                 (\bm{P}^{new}-\bm{\mathcal{Q}}\hat{\bm{w}})] \;.
\end{equation}
The expectation is taken over $\bm{P}^{new}$. In the following we will denote
the predictor $\bm{\mathcal{Q}{\hat w}}$\/ as $\hat {\bm P}$\/ and call it
the fitting histogram.

Following reference \cite{breiman}, $V$-fold Cross-Validation allows
to estimate\\ $PE(\bm{\mathcal{Q}{\hat w}})$. Here the given data set
$\mathcal U$\/ is split into $V$\/ subsets $\mathcal U_1,...,\mathcal U_V$\/
with equal number of events. The complementary sets are denoted by
$\mathcal U^{(v)}=\mathcal U-\mathcal U_v$. Applying the minimisation
procedure to $\mathcal U^{(v)}$ and forming the predictors
$\bm{\mathcal{Q}}\hat{\bm{w}}^{(v)}$, the Cross-Validation
error ($CV$) is defined by
\begin{equation}
 CV = \frac{1}{n}\sum_{v=1}^V
         (\bm{P}_{v}-\bm{\mathcal{Q}}\hat{\bm{w}}^{(v)})^t
          \bm{\mathrm{C}}^{-1}
         (\bm{P}_{v}-\bm{\mathcal{Q}}\hat{\bm{w}}^{(v)}) \;,
\end{equation}
where $\bm{P}_{v}$\/ is the vector of histogram contents for the subset of
the data $\mathcal U_v$. The Cross-Validation error is the estimate of
the Prediction Error
\begin{equation}
  CV=\widehat{PE}(\bm{\mathcal{Q}{\hat w}}) \;.
\end{equation}
In order to have sufficient sampling of the configuration space, the number
of folders used in the Cross-Validation procedure should not be too small.
On the other hand, for statistically meaningful results, it should
not be too large either. Practice shows that taking $V$\/ in the range between
5 and 10 usually gives satisfactory results, and that the performance
is not sensitive to the exact choice.

\vspace*{5mm}
\noindent
The proposed unfolding procedure consists of three steps:

\subsection{First step}
The positions $\{x_i\}$\/ of  the PDFMs are drawn randomly from a uniform
distribution on the allowed range of $x$\/ and with a number of PDFMs such
that the average distance between individual centers is significantly
smaller than the width of the PDFMs.

In order to minimise the loss of information due to binning, the number of
bins for the measured histogram $\bm{P}$\/ should be as large as possible.
On the other hand, in order to have meaningful error estimates for the
least squares fits that determine $\bm{\hat{w}}$, the number of entries in
a single bin should not be less than $25$. Binning with approximately
equal number of events in each bin is preferable.

In this first step the width for all PDFMs in the mixture is taken to be the
same, $\lambda_i=\lambda$. Different values $\lambda$\/ are tried and the
$\hat \lambda$\/ with the smallest Cross-Validation error
\begin{equation}
 \hat{\lambda}= \underset{\lambda} {\mathrm{argmin}} ~CV(\lambda)
\end{equation}
is selected.

\subsection{Second step}
Exploiting the information gained so far, the procedure is repeated with
positions $\{x_i\}$\/ of the PDFMs randomly drawn according to the estimate
of the true density $\hat p(x)$ (\ref{px}) obtained in the first step.
In addition, the widths of the PDFMs  $\{\hat \lambda_i\}$\/ are taken to
be inversely proportional to the square root of the result from the first
step at the position $\{x_i\}$
\begin{equation}
\hat \lambda_i= \frac{\hat \lambda}{\sqrt{\hat{p}(x_i)}} \label{lambda}
\end{equation}
with the value of $\hat\lambda$\/ again determined by means of Cross-Validation.
This second step is motivated by the results of reference \cite{abramson},
where it is shown that by this way the bias of a kernel estimation of a PDF can be
decreased. The approach balances better smoothing in less densely populated
regions against the possibility to resolve finer structures in regions
with a higher sampling. It is plausible that  for the unfolding case
the bias on the shape of the density estimate will decrease also. Finally it
has to be noted that this second step can be iterated several times, even
though practical examples show that the gain is small.

It is recommended to use the same number of PDFMs for the second step as in the first step or  more.

\subsection{Third step}
Since the number of terms obtained by the previous two steps can still be
large, with not all PDFMs contributing independent information, a third step is added to select the most relevant subset. To reduce the number of the PDFMs, the non-negative garrote method \cite{breiman}  is used. It
amounts to taking the set of non-zero weights $\{\hat w_j\}$\/ obtained
in the second step and finding coefficients $\{c_j\}$\/ that minimise
\begin{equation}
\sum_{i=1}^n(P_i-\sum_{j=1}^s  \mathcal Q_{ij}c_j{\hat w_j})^2/\sigma^2_i
\end{equation}
under the constraints
\begin{equation}
c_j \geq 0 \quad\mbox{and}\quad \sum_{j=1}^s c_j \leq r \;.
\end{equation}
For $r \geq \sum \hat w_i$\/ the solution from the previous step is not
touched. For smaller values the garrote eliminates some of weights and
modifies others, such that $\tilde{w}_j(r)=c_j {\hat w_j}$\/ are the new
values of the weights for the PDFMs of the estimate the unfolded distribution.
Cross-validation is used to choose the optimal garrote parameter $r$.
To reduce a potential bias introduced by the garrote, the weights of the PDFMs are again determined by a non-negative least squares fit
on the remaining terms.

\subsection{Quality assessment and error propagation}
The quality of the fit can be assessed with common tools used in regression
analysis \cite{seber}:
\begin{enumerate}
  \item $p$-value of the fit is defined by $Pr(X \geq \sum_{i=1}^n(P_i-\hat P_i)^2/\sigma_i^2)|\chi_{n-s}^2)$, where $Pr$ stands for
      probability
  \item analysis of the normalised residuals $Res_i= (P_i-\hat P_i)/\sigma_i, i=1,...,n$
  \begin{enumerate}
      \item as a function of the estimated value $\hat{\bm P}$
      \item as a function of the observed value  $x'$
  \end{enumerate}
  \item Q-Q plot:  (data quantile)$_i$=
   (number of residuals $\leq Res_i)/n$
    versus\\
    (theoretical quantile)$_i = Pr(X \leq$ ($Res_i |\mathcal{N}(0,1)$),  $i=1,...,n$
\end{enumerate}

Since the unfolding procedure described above is not analytically defined,
the bootstrap approach \cite{boot} is the method of choice to estimate the
statistical uncertainties of the unfolding result. Keeping the normalisation
of the observed histogram constant, replications are generated according to
the multinomial distribution
\begin{equation}
\frac{N!}{N_1!N_2! \ldots N_n!}\mathcal P_1^{N_1} \ldots \mathcal P_n^{N_n}
\quad\mbox{with}\quad
\mathcal P_i = \frac{\sum_{j=1}^s \mathcal Q_{ij}{\hat w_j}}
                    {\sum_{i=1}^n\sum_{j=1}^s \mathcal Q_{ij}{\hat w_j}}\;,
\end{equation}
where the set of positive weights obtained in the final step is used.

A histogram representation $\hat{\bm{p}}$\/ for the unfolded distribution
$\hat{p}(x)$\/ with $m$\/ bins integrating over the $x$-intervals
$[b_{i-1},b_i],\,i=1,\ldots,m$\/ is obtained by
\begin{equation}
  \hat{\bm{p}}=\bm{\mathrm K} \, \bm{\hat w} \;,
\label{unfbin}
\end{equation}
where $\bm{\mathrm K}$\/ is an $m\times k)$\/ matrix with elements
\begin{equation}
   \mathrm{K}_{ij} = \int_{b_{i-1}}^{b_i} K_j(x ; x_j,\lambda_j)\, dx \;.
\end{equation}

The unfolding method described above assumes that the matrix
$\bm{\mathrm{Q}}$\/ relating the weight vector $\bm{\hat{w}}$\/ to the
measurements $\bm{P}$\/ is known exactly. Therefore, when $\bm{\mathrm{Q}}$\/
is determined by means of a Monte Carlo simulation, the Monte Carlo sample
should be significantly larger than the data sample.

In an extension of the method which is applicable also in cases where the
Monte Carlo statistics is of the same order or less than the data statistics
is obtained by using a modified matrix of errors $\bm{\mathrm{C}}$\/ which
includes statistical errors for the elements of matrix $\bm{\mathrm{Q}}$\/
\cite{fedorov}, and the Cross-Validation statistics substituted by the
goodness-of-fit statistics for the comparing unweighted $\bm{P}_{v}$\/ and
weighted $\bm{\mathcal{Q}}\hat{\bm{w}}^{(v)}$\/ histograms given in
references \cite{gagunashviliph, gagunashvilipro}.

\section{Numerical examples}
Three types of numerical examples are discussed to illustrate the unfolding
procedure. The first  is the classic example of a double peak structure
proposed by V.~Blobel \cite{blobel}.  The second is a strongly
varying one-sided distribution, and the third one a two-dimensional case.

\subsection{Double peak structure}
\label{sec:example1}
\begin{figure}[H]
\begin{center}
\begin{tabular}{cc}
\hspace *{-0.8 cm}
\subfigure{\includegraphics[height=0.41\textwidth]{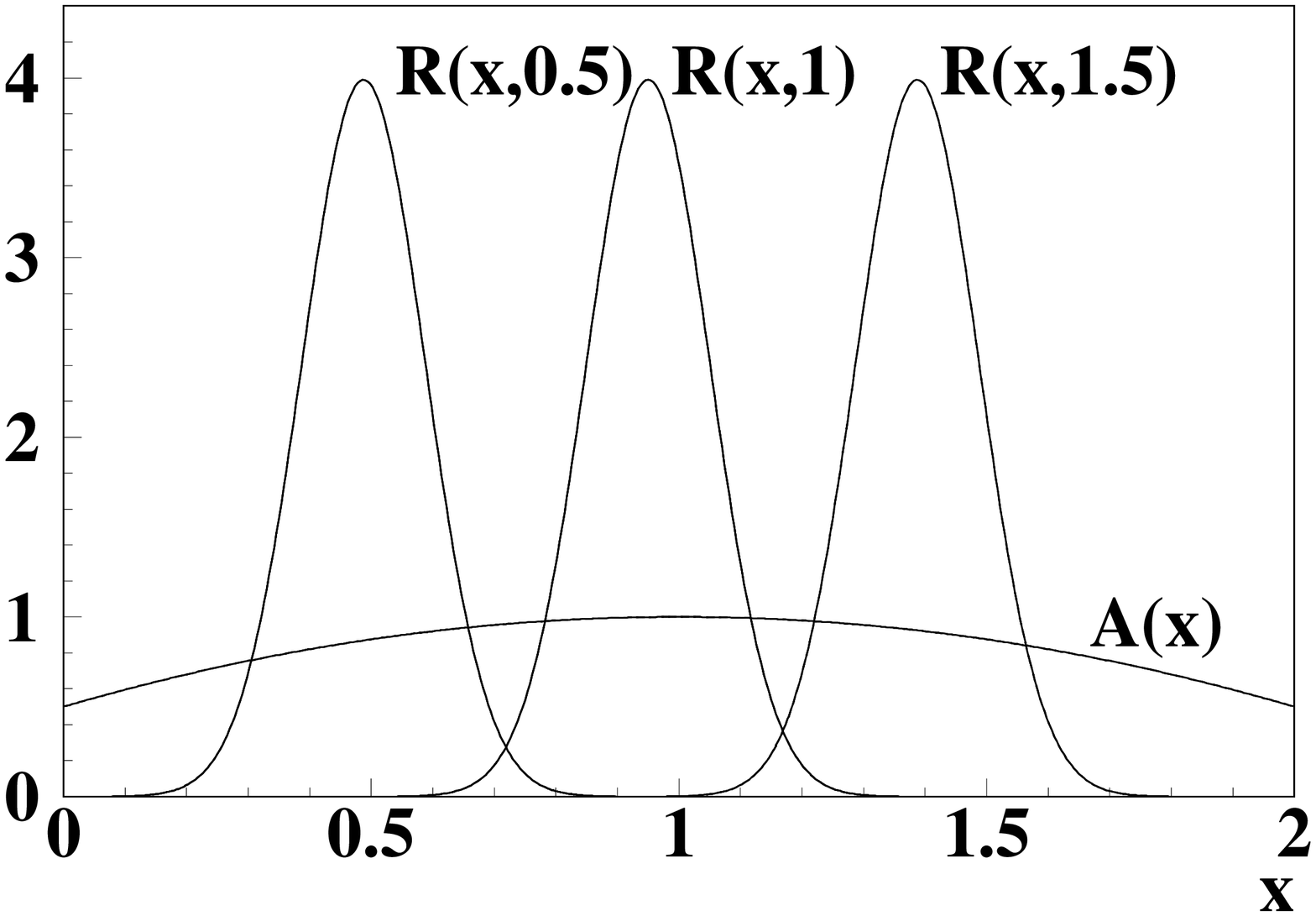}} & \hspace *{-0.8 cm}
\subfigure{\raisebox{0mm}{\includegraphics[height=0.41\textwidth]{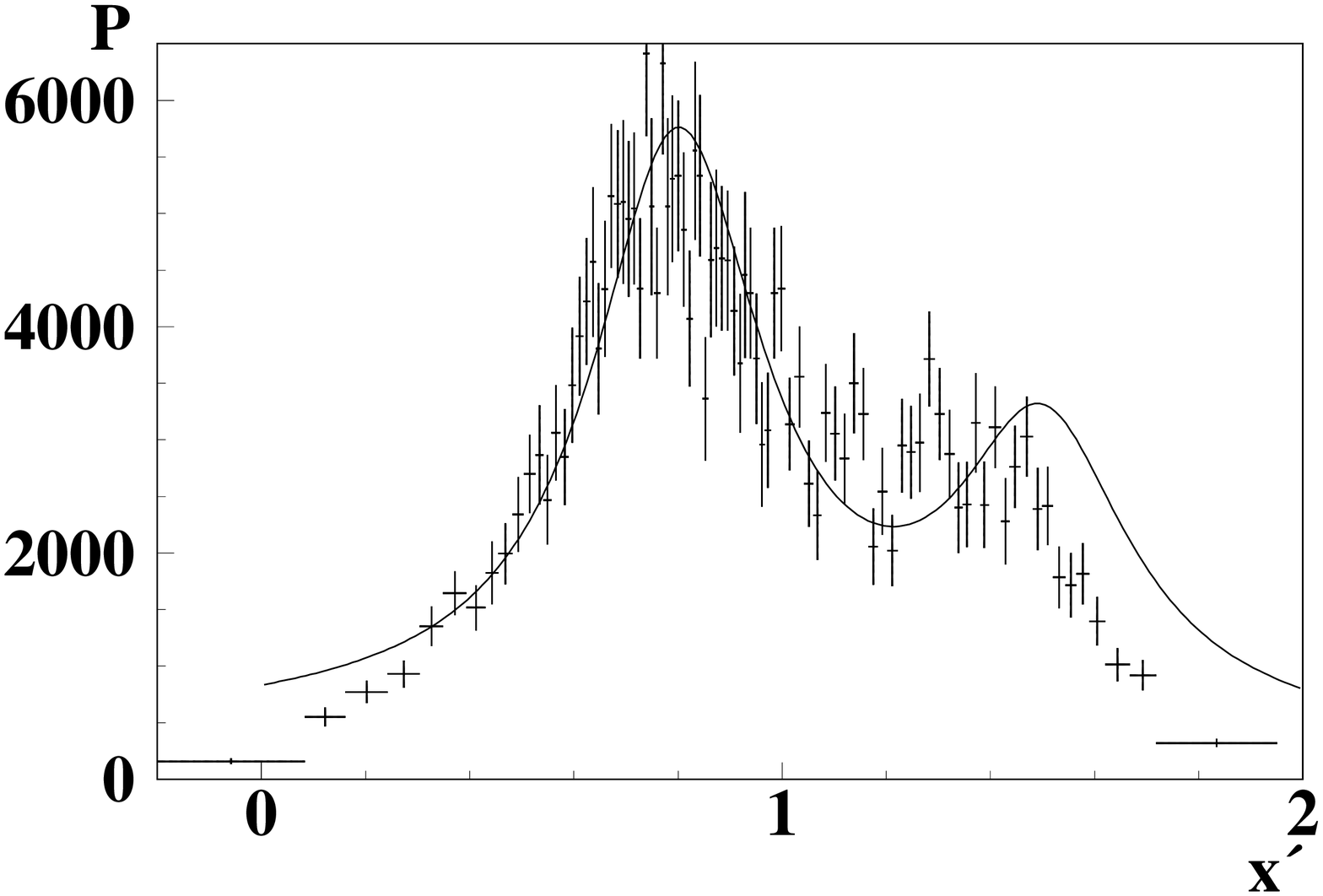}}}
\end{tabular}
\end{center}
\vspace *{-1cm}
\caption{Acceptance function $A(x)$\/ and resolution function $R(x'|x)$\/
         for $x=0.5, 1.0$ and $1.5$ (left) and histogram of the measured
         distribution $\bm{P}$\/ based on a sample of 5\,000 events generated
         for the true distribution (right). The bin contents of the
         histogram are normalised to the bin width. The true distribution $p(x)$\/ is shown  by the curve (right).}
\label{fig:resaccept}
\end{figure}

The method described above is illustrated using the example proposed in
reference \cite {blobel}. The true distribution, defined on the range
$x \in [0,2]$\/ is described by a sum of three Breit-Wigner functions
\begin{equation}
p(x) \propto \frac{4}{(x-0.4)^2+4}
       +     \frac{0.4}{(x-0.8)^2+0.04}
       +     \frac{0.2}{(x-1.5)^2+0.04}
\label{testform}
\end{equation}
from which the experimentally measured distribution is obtained by
\begin{equation}
P(x') \propto \int_0^{2} p(x)A(x)R(x'|x)dx ,
\end{equation}
with an acceptance function $A(x)$
\begin{equation}
A(x)=1-\frac{(x-1)^2}{2}
\end{equation}
and a response function describing a biased measurement with gaussian smearing
\begin{equation}
R(x'|x)=\frac{1}{\sqrt{2\pi}\sigma}\exp\left(-\frac{(x'-x+0.05x^2)^2}{2\sigma^2}\right)
\quad\mbox{with}\quad
\sigma=0.1 \;.
\end{equation}
The acceptance and resolution functions are shown in Fig.~\ref{fig:resaccept}.
Also shown is an example for the measured distribution obtained by simulating
a sample of $N=5\,000$\/ events. A histogram with number of bins $n=87$\/ and
approximately equal number of events in each bin was used.
\vspace *{-1. cm}

The PDFMs were defined in the form
\begin{align*}
 K_i(x;x_{i},\lambda_i)\propto
 \bigg[
   \frac{1}{\lambda_i \sqrt{2\pi}}\exp\left({-\frac{(x-x_{i})^2}{2\lambda^2_i}}\right)
  +\frac{1}{\lambda_i \sqrt{2\pi}}\exp\left({-\frac{(x+x_{i})^2}{2\lambda^2_i}}\right)\\
  +\frac{1}{\lambda_i \sqrt{2 \pi}}\exp\left({-\frac{(x-4+x_{i})^2}{2\lambda^2_i}}\right)
 \bigg] \, I_{\{x\in[0;2]\}},
\end{align*}
with the indicator function
\begin{equation*}
    I_{\{\cdots\}} = \left\{ \begin{array}{l}
                            1 \qquad \mbox{if the condition given in the curly
                                           brackets is satisfied}\\
                            0 \qquad \mbox{otherwise}
                           \end{array}\right.  \;.
\end{equation*}
The functional form is chosen in accordance with the recommendation formulated in
reference \cite{silverman} for PDFs defined on the restricted interval. An
initial set of 400 PDFMs was used with positions $x_{i}$\/ uniformly distributed
over the interval $[0,2]$. For the determination of the matrix
$\bm{\mathrm{Q}}$\/ a sample of 500\,000 Monte Carlo events was simulated.
Here  a uniform true distribution was taken and the responses of the
individual PDFMs were calculated by weighting the Monte Carlo
events with weights proportional to the value of the respective PDFM \cite{sobol}.
\vspace *{-1cm}
\begin{figure}[H]
\centering
\includegraphics[width=0.9\textwidth]{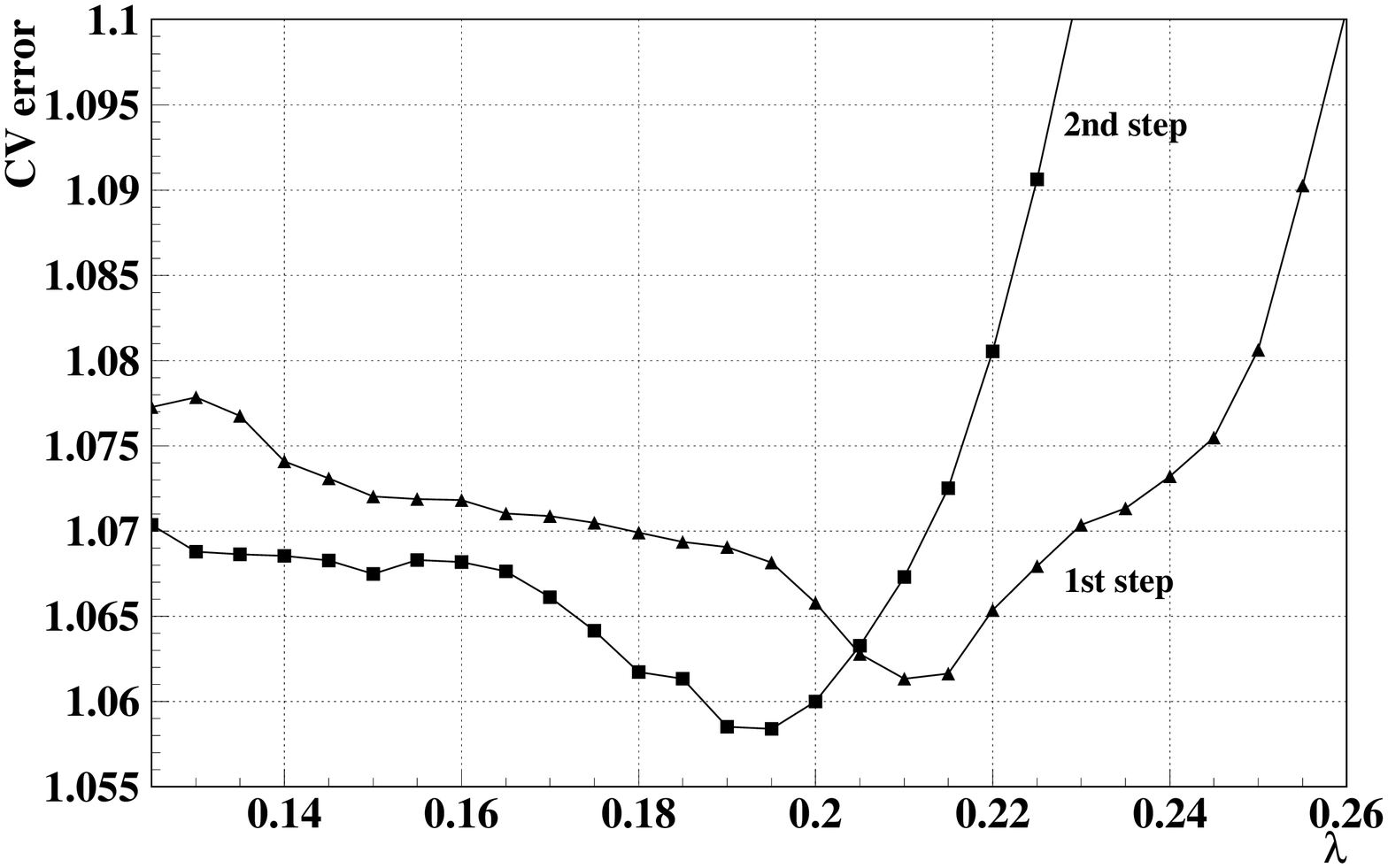}\\ \vspace *{-1.5cm}
\includegraphics [width=0.9\textwidth]{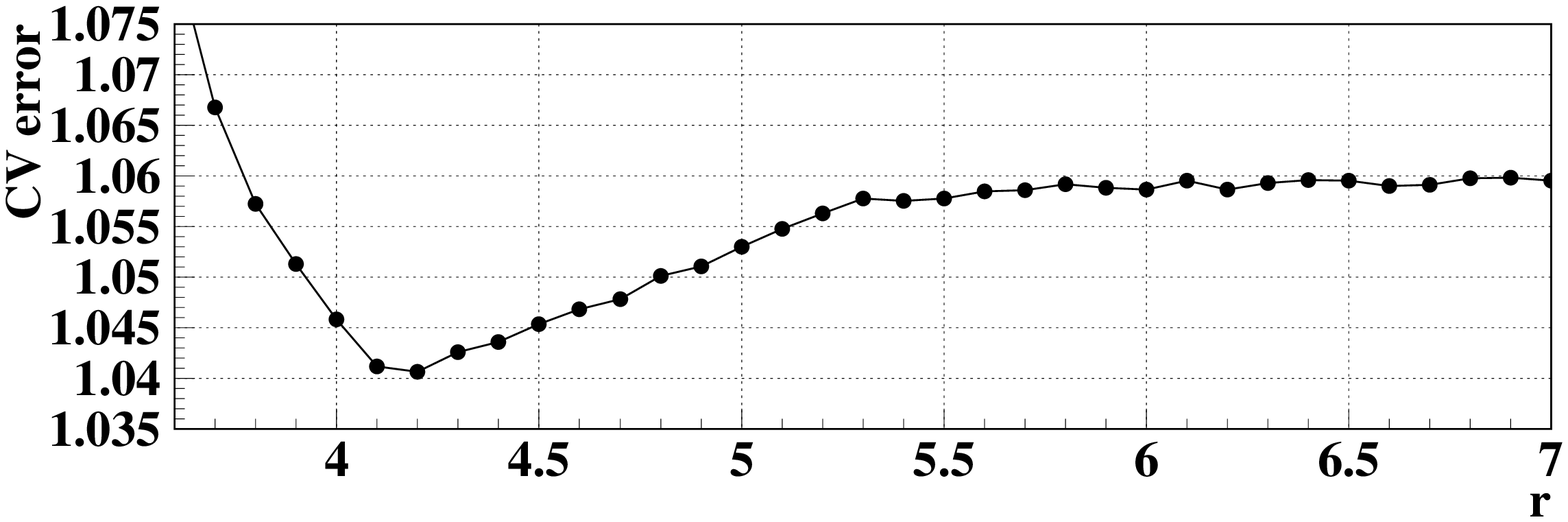}
\vspace *{-1. cm}
\caption{Cross-Validation errors of the first and second
         unfolding step as a function of $\lambda$\/ (top),
         and Cross-Validation error in the third step for
         $\hat\lambda=0.19$\/ as a function of the
         garrote parameter $r$ (bottom).}
\label{fig:cross1}
\end{figure}

The top plot of Fig.\,\ref{fig:cross1} shows how the Cross-Validation
error for 5-fold Cross-Validation in the first and the second step behaves
as a function of $\lambda$. One observes that the best value in the second
step is slightly smaller in step 2, and also that the Cross-Validation error
is reduced. This shows that adapting the widths of the PDFMs in the model
according to the estimated density not only provides better smoothing
in regions of small statistics, but also improves the quality of the
unfolding result. The best value is found as $\hat\lambda=0.19$. This value
is used in the third step, where the non-negative garrote is employed to
reduce the number of the PDFMs. The bottom frame of
Fig.\,\ref{fig:cross1} shows the Cross-Validation error as a function
of the garrote parameter $r$. Again a significant improvement is found
for $r=4.2$. For the statistics used in this example, the final estimate
of the true distribution contains only three terms.

\begin{figure}[H]
\begin{center}$
\begin{array}{cc}
\includegraphics[width=9cm]{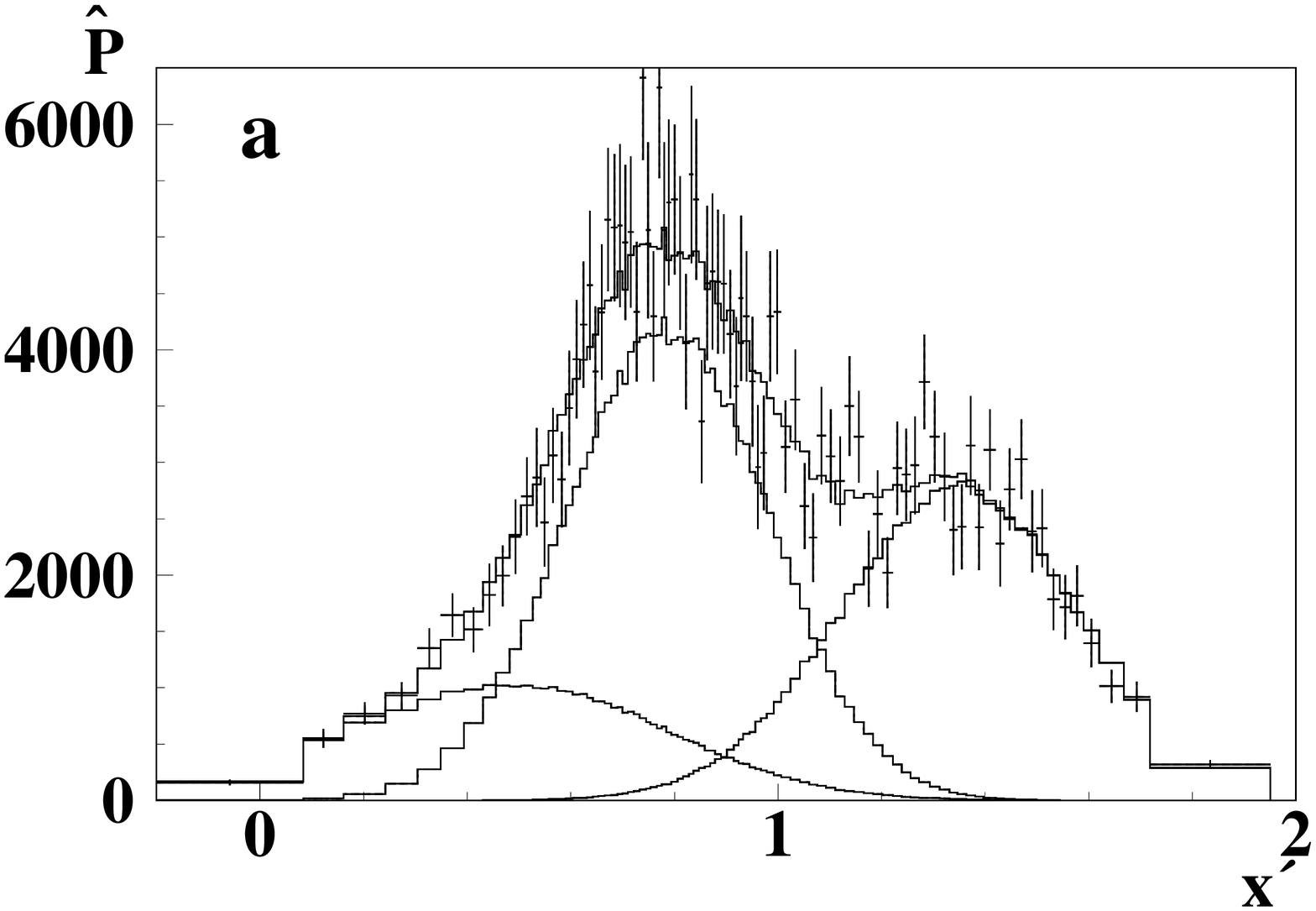} &
\vspace *{-1.1 cm} \hspace *{-0.6cm}\includegraphics[width=5.3cm]{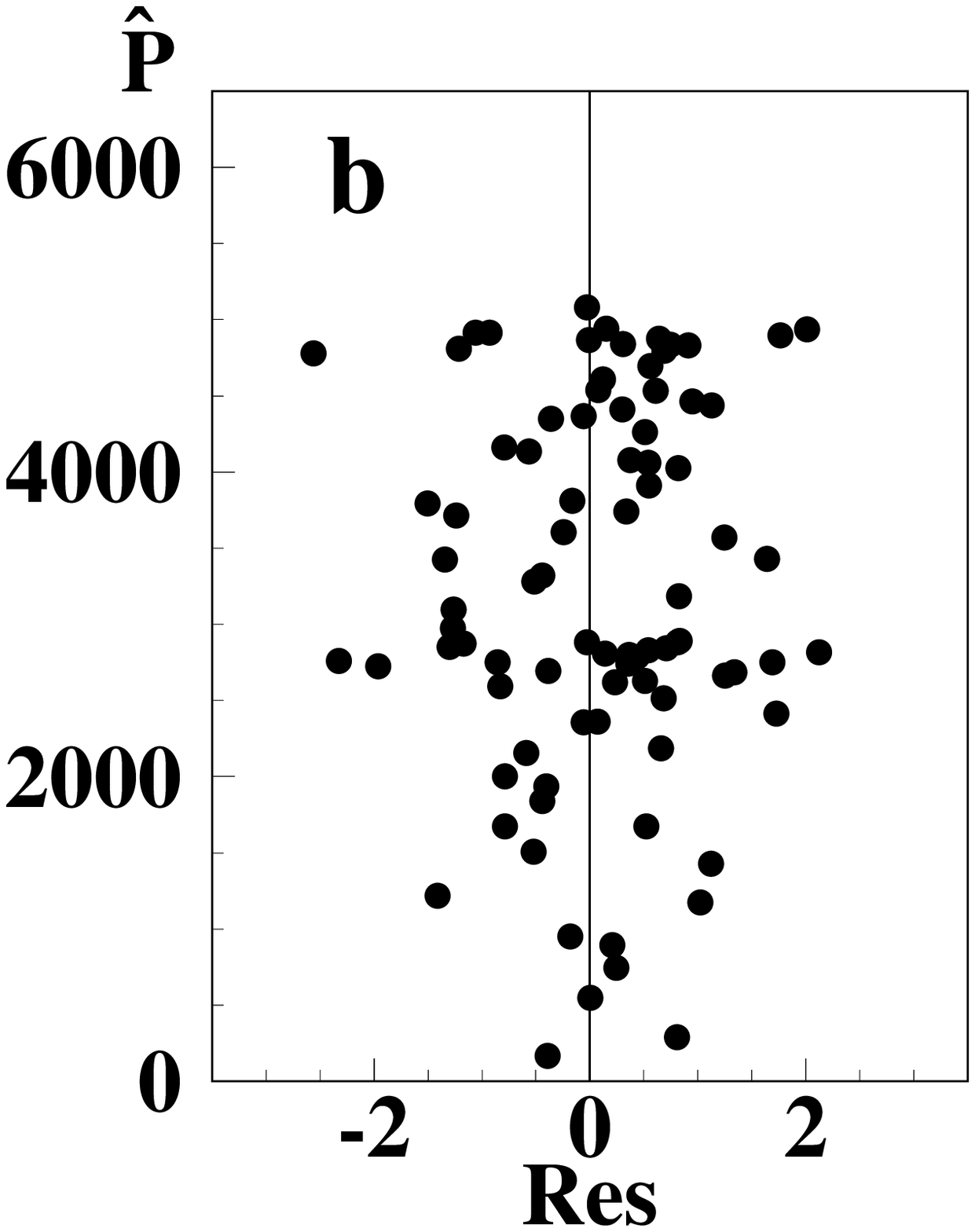} \vspace *{0.3cm}\\
\includegraphics [width=9cm]{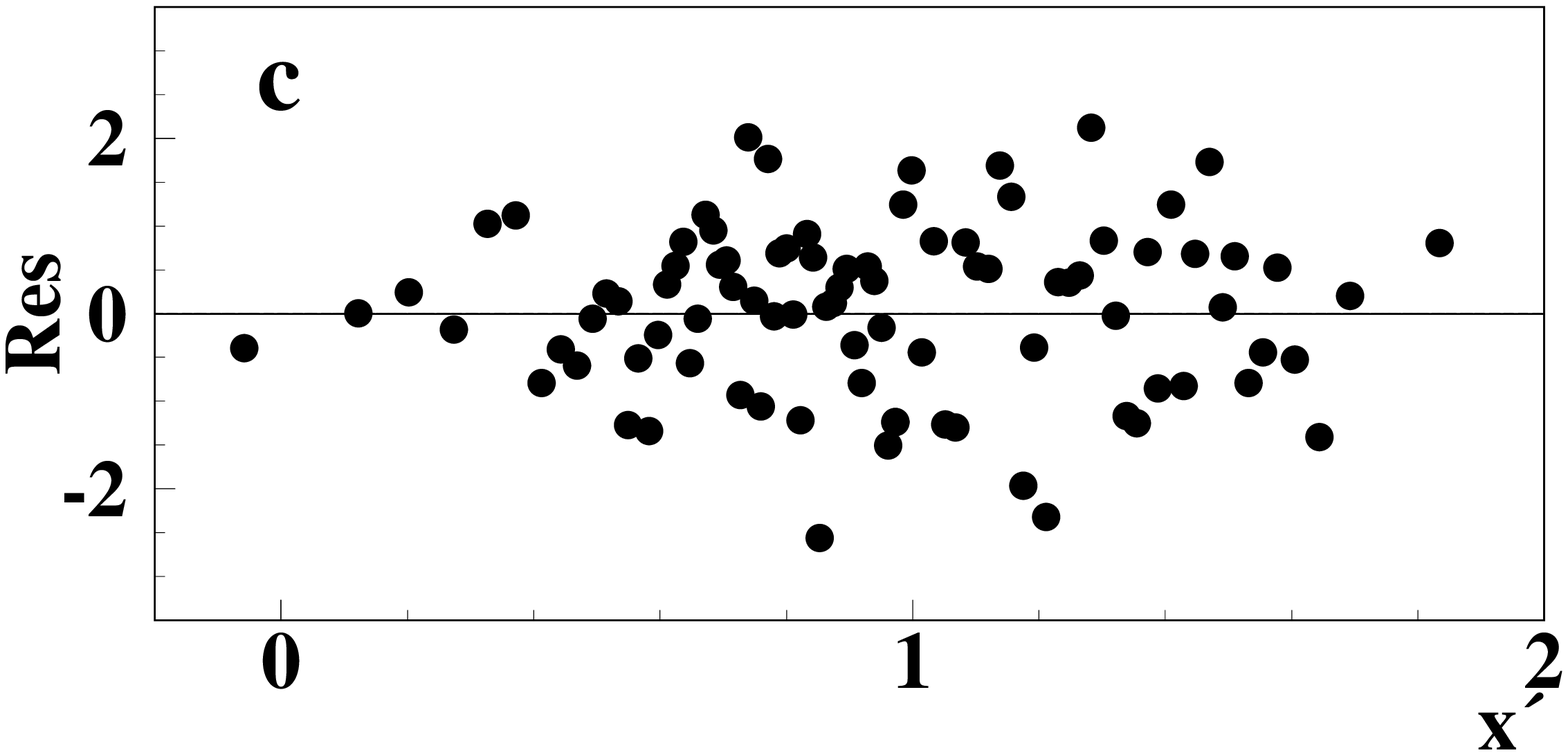}&
\vspace *{-1.1 cm} \hspace *{-0.6cm}\includegraphics[width=5.3cm]{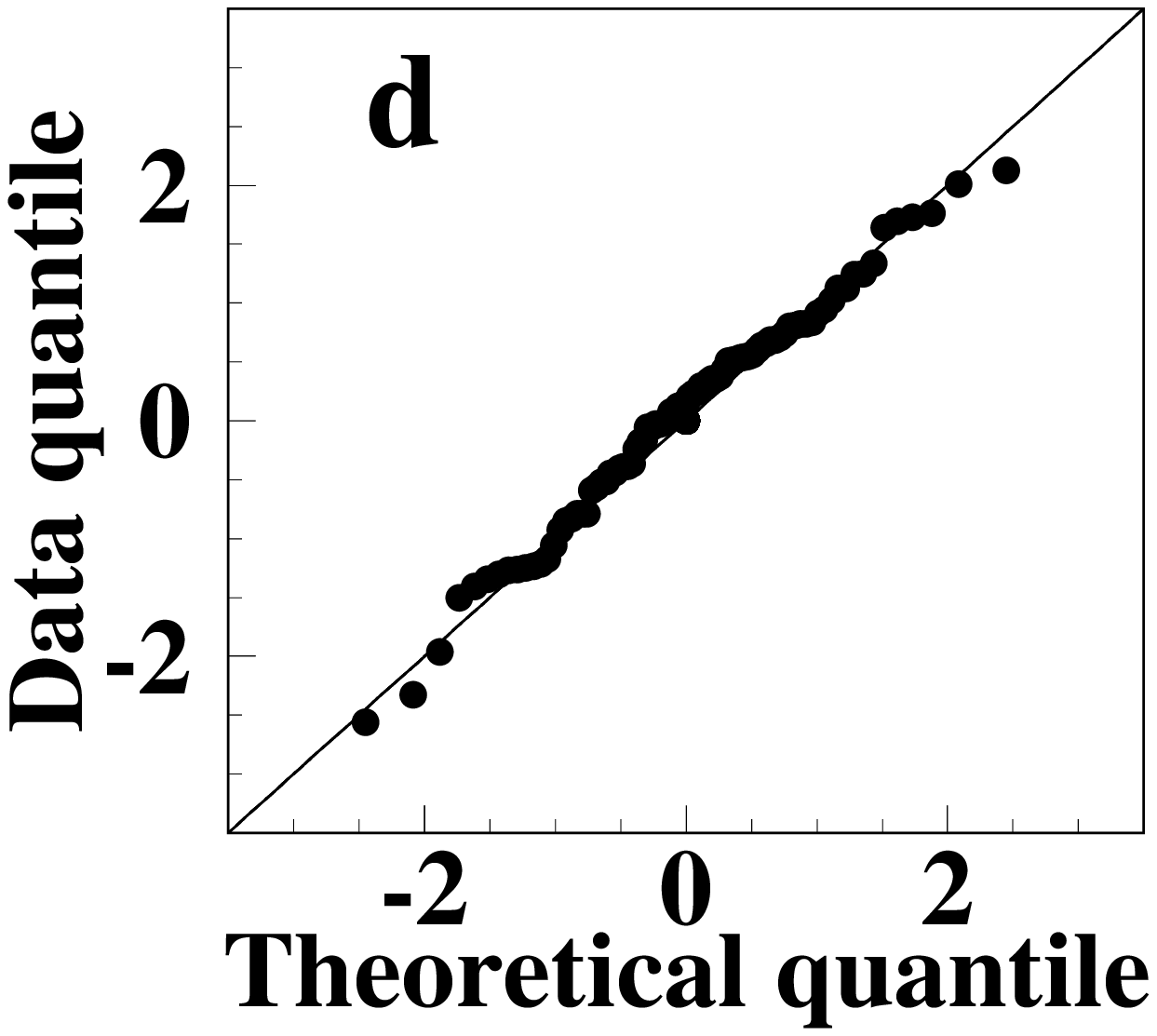}
\end{array}$
\end{center}
\vspace *{0.5cm}
\caption{Illustration of the quality of the unfolding result.
        (a) folded  estimate of the true distribution $\hat{\bm{P}}$ (solid histogram), with tree components,
        compared to the measured distribution $\bm{P}$; (b) normalised residuals
        of the fit as a function of $\hat{\bm{P}}$; (c) normalized
        residuals as a function of $x'$; (d) quantile-quantile plot
        for the normalized residuals.}
\label{fig:quality1}
\end{figure}

The quality of the unfolding result is illustrated by Fig.~\ref{fig:quality1}.
It shows how the folded estimate of true distribution  $\hat{\bm{P}}$\/, with three components, approximates the
measured distribution, together with plots of the residuals and the
quantile-quantile plot. No structure in either of the control plots
is observed. The $p$-value from the test comparing the histogram of the
measured distribution $\bm{P}$\/ and the fitting histogram $\hat{\bm{P}}$\/ is $p=0.6$.
\vspace *{-0.7 cm}

\begin{figure}[H]
\begin{center}$
\begin{array}{cc}
\vspace*{-1.92cm}\includegraphics[width=2.99in]{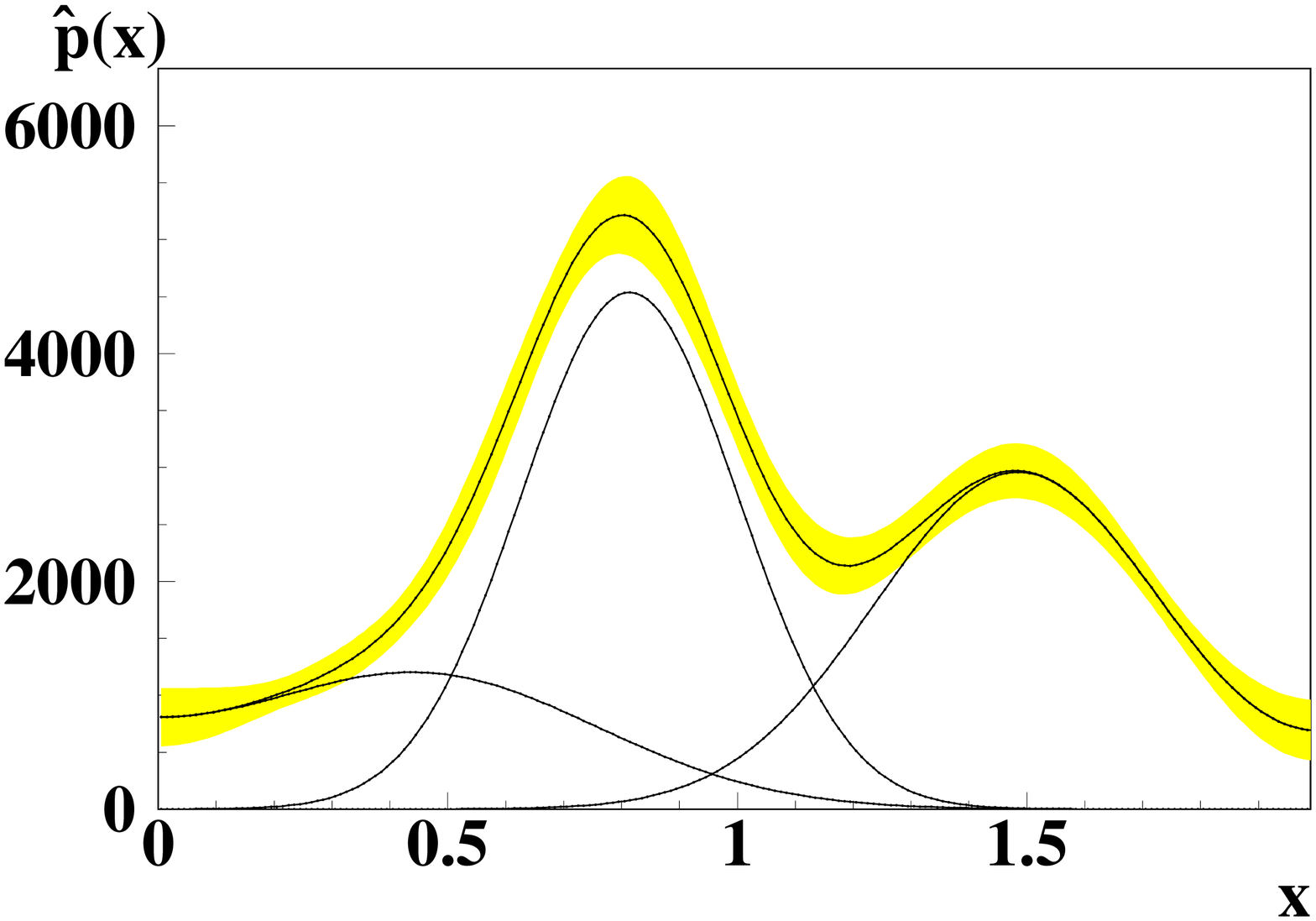} &
\hspace*{-1.82cm}\includegraphics[width=2.99in]{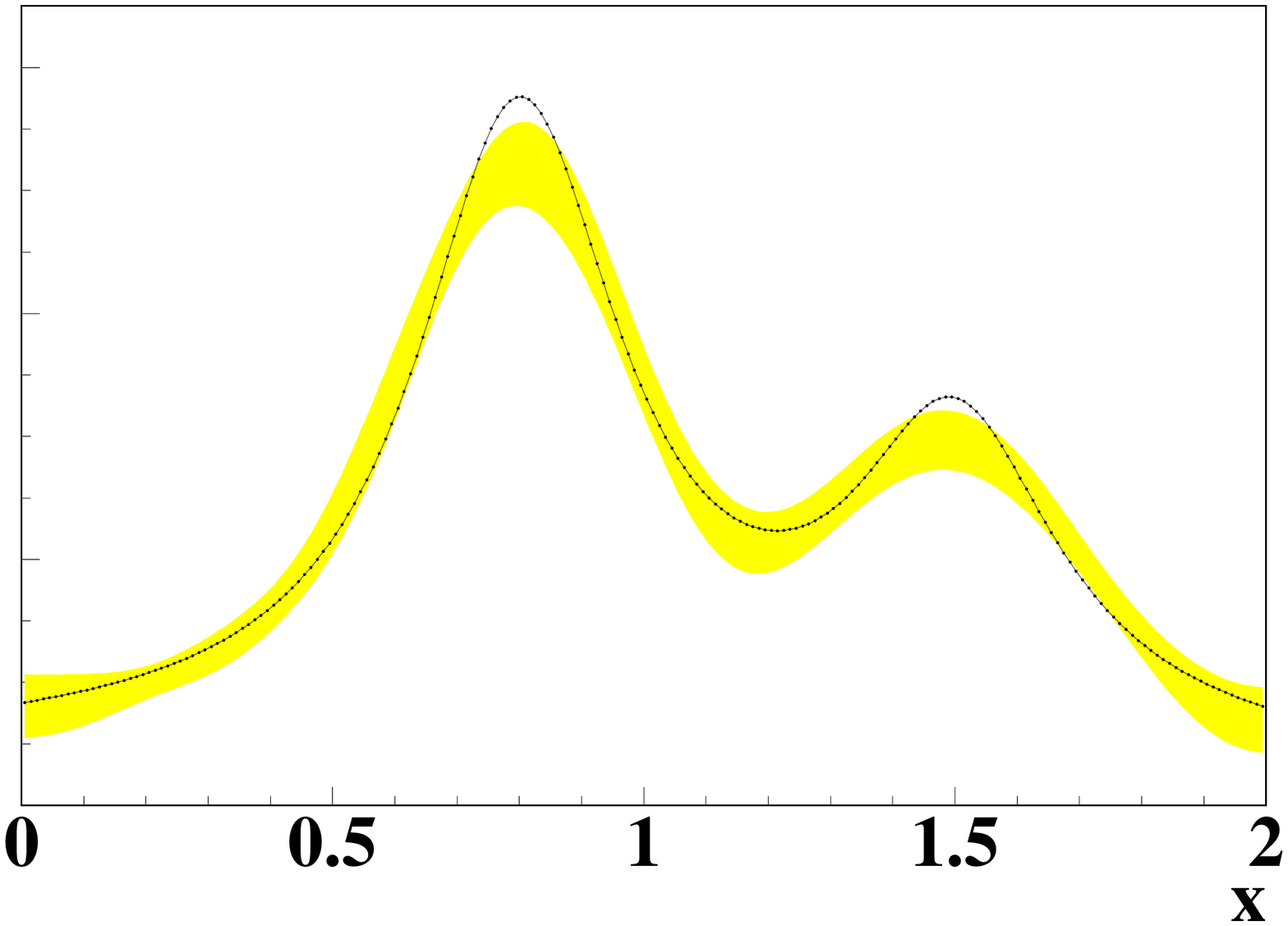}\\
\end{array}$
\end{center}
\vspace*{1.0cm}
\caption{Components of the unfolded distribution and the unfolded
         distribution $\hat{p}(x)$\/ given by the sum of the components
         with $\pm 2\delta(x)$\/ interval (left) and the two standard deviation band overlaid with the true distribution $p(x)$\/ (right).}
\label{fig:unfolded1}
\end{figure}

\begin{figure}[H]
\begin{center}$
\begin{array}{cc}
\vspace*{-1.92cm}\includegraphics[width=2.99in]{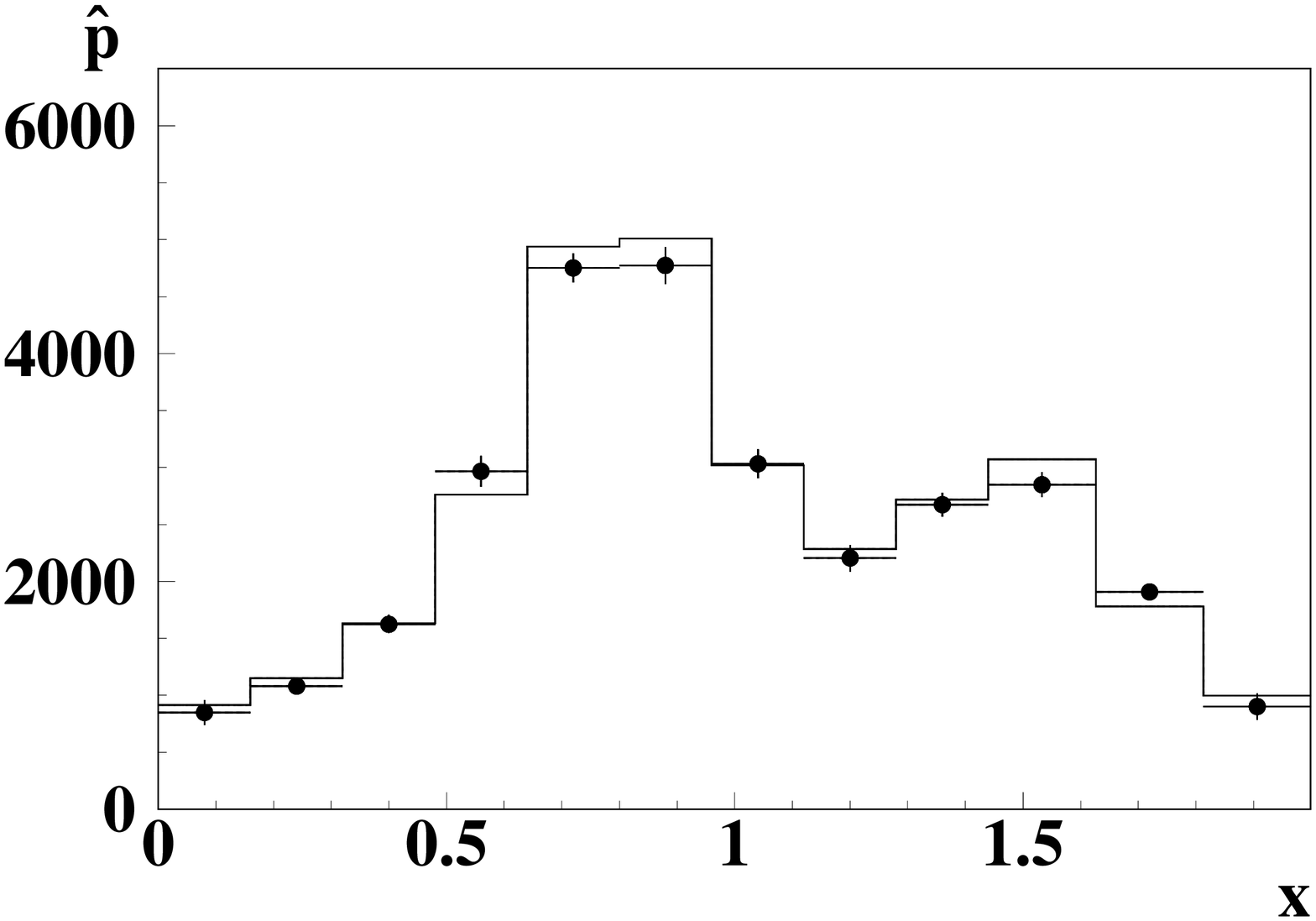} &
\hspace*{-1.82cm}\includegraphics[width=2.99in]{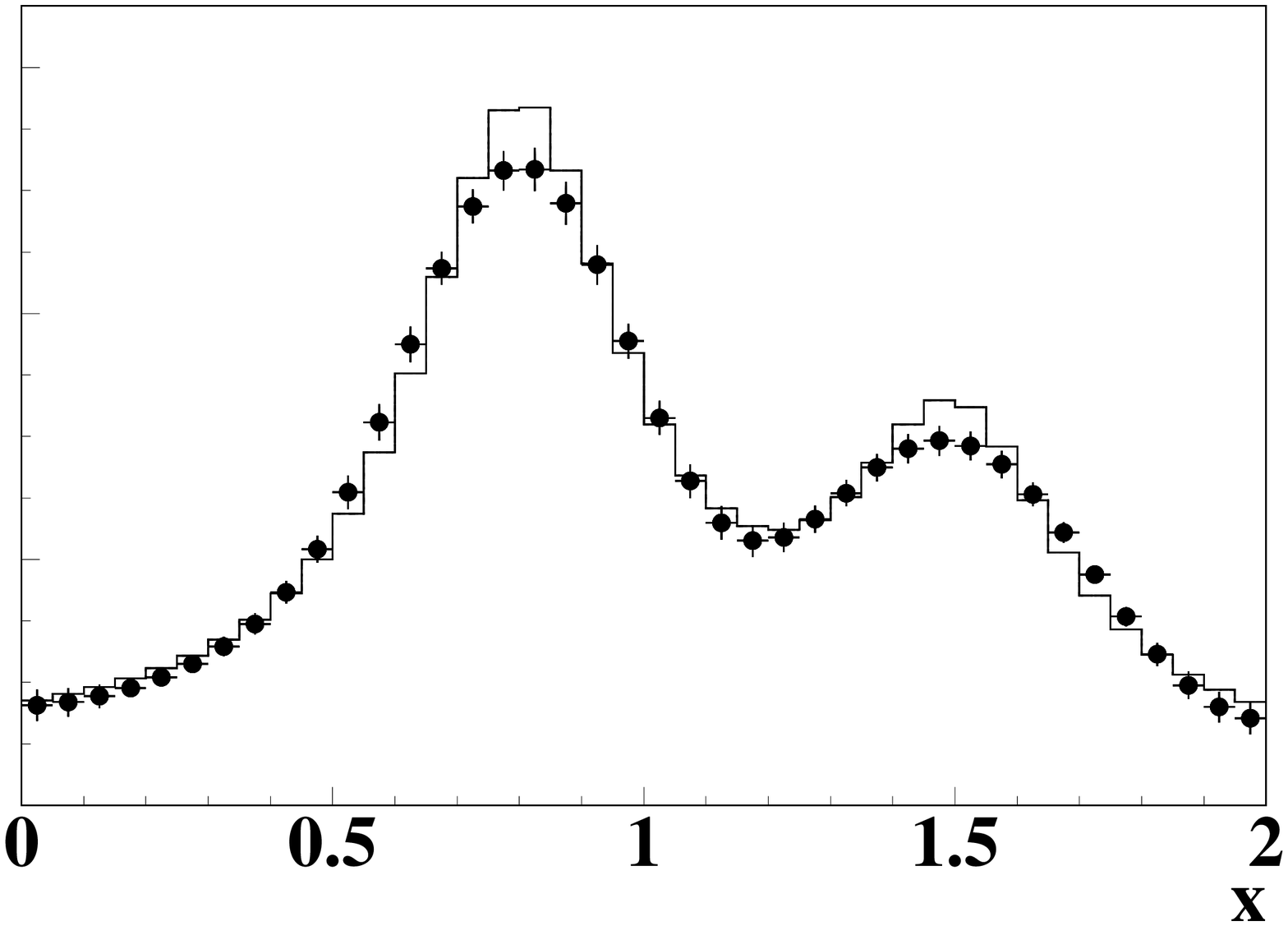}\\
\end{array}$
\end{center}
\vspace*{1.0cm}
\caption{Binned representation of the unfolding result $\hat{p}_i$\/ for
         $n=12$ (left) and $n=40$ (right) bins. The points with error
         bars are the estimate obtained by the unfolding procedure,
         the histogram shows the true bin contents $p_i$.}
\label{fig:unfoldedhist1}
\end{figure}

The estimate of the true distribution obtained by the unfolding procedure
is presented in Figs.\,\ref{fig:unfolded1} and \ref{fig:unfoldedhist1}.
The components of the unfolding results
together with the estimate $\hat{p}(x)$ in Fig.~\ref{fig:unfolded1}.
Also shown are the two standard deviation bands $\pm 2\delta(x)$\/ compared
to the true distribution $p(x)$. Histogram presentations of the unfolded
distribution are shown in Fig.\,\ref{fig:unfoldedhist1} for $n=12$\/ bins
as in reference \cite{blobel} and for $n=40$\/ bins as in reference \cite{hocker}.
Standard deviation bands and bin-by-bin uncertainties for the histograms
were estimated by the bootstrap method.
\vspace *{-1. cm}

\begin{figure}[H]
\begin{center}$
\begin{array}{cc}
\vspace*{-1.92cm}\includegraphics[width=2.99in]{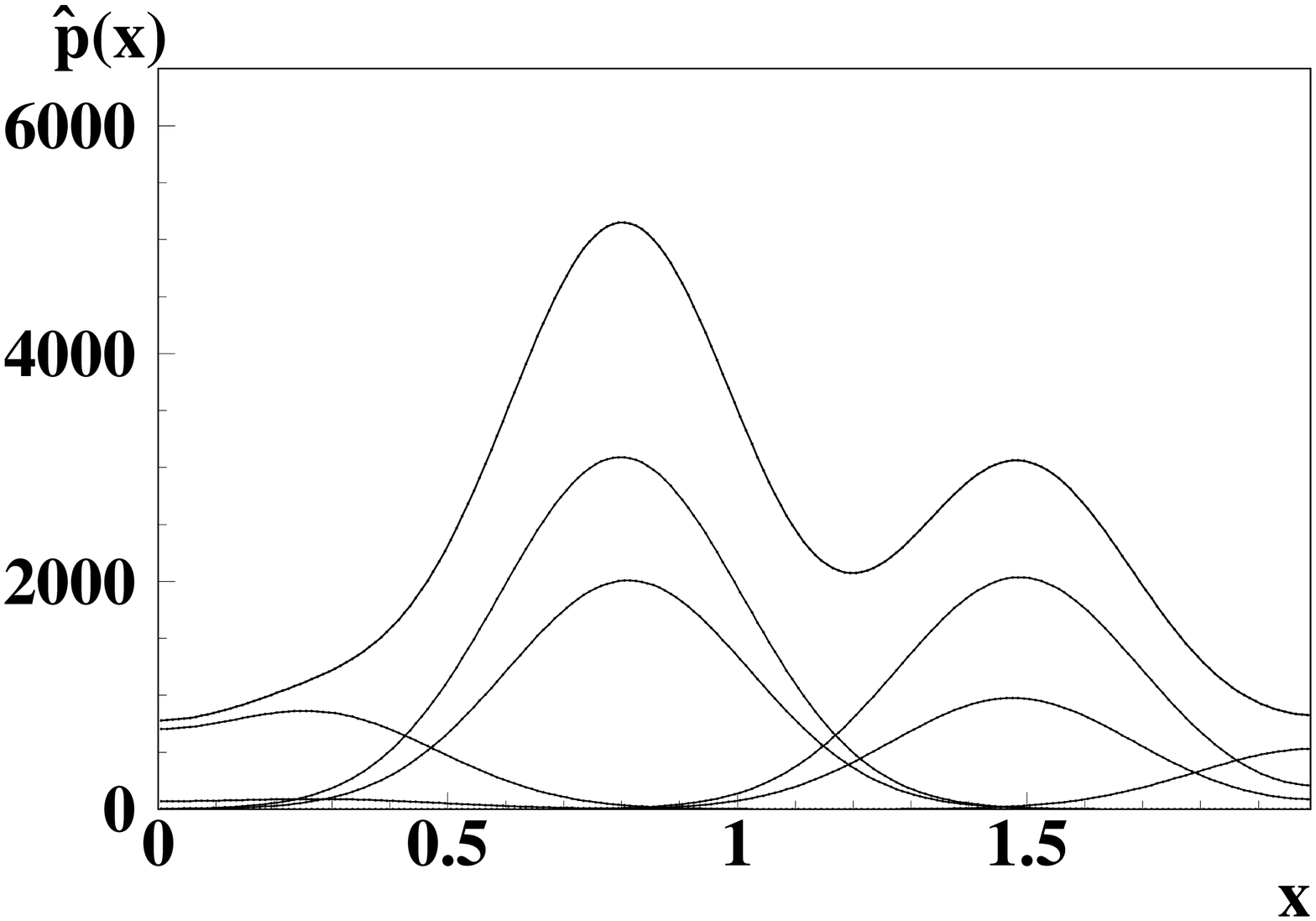} &
\hspace*{-1.82cm}\includegraphics[width=2.99in]{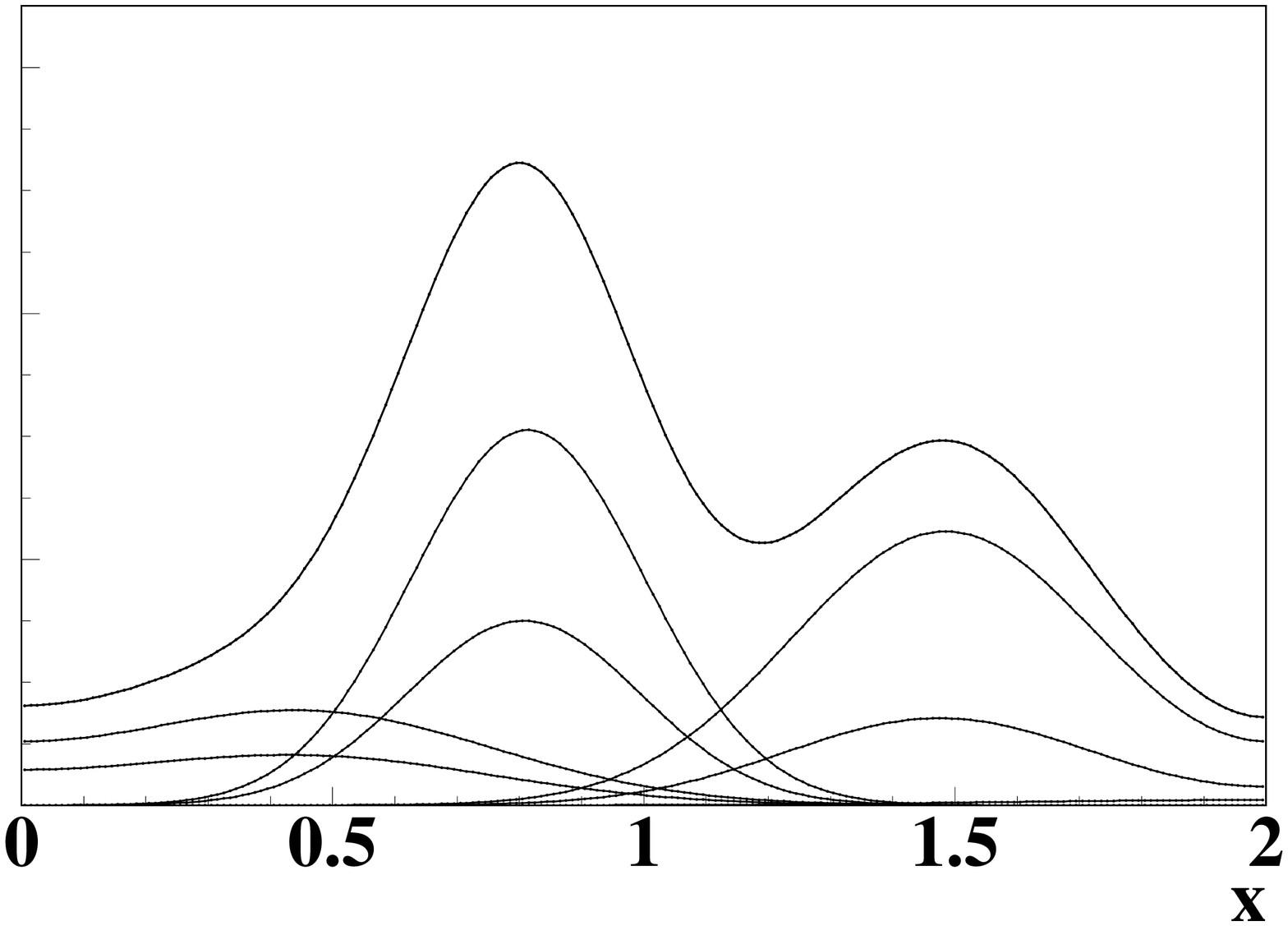}\\
\end{array}$
\end{center}
\vspace*{1.0cm}
\caption{Components of the unfolded distribution and the unfolded
         distribution $\hat{p}(x)$\/ given by the sum of the components
         for the step 1 (left) and for the step 2 (right).}
\label{fig:unfolded1122}
\end{figure}

For the illustration of the proposed algorithm,  the  unfolded distributions    for steps 1 and  2  on Fig.~\ref{fig:unfolded1122}  are presented  as well as  Table 1 with values of parameters $x_i$, $\hat \lambda_i$, $\hat w_i$ of the components for the three steps of the procedure.

 \begin{table}[htb]
\centering
\caption{Values of  parameters $x_i$, $\hat \lambda_i$, $\hat w_i$ of the   PDFMs  represented unfolded distributions for  three steps of procedure}
\label{tab:xval}
\begin{tabular}{c|ccccccc|c}

 $i$   & 1  & 2 & 3  & 4 & 5 & 6 & 7  \\
\hline
 $x_i$   & 0.275 & 1.475  & 1.992 & 1.487  & 0.798 & 0.812  & 0.284 &  \\
 $\hat \lambda_i$ & 0.210 & 0.210  & 0.210 & 0.210  & 0.210 & 0.210  & 0.210 & Step 1 \\
 $\hat w_i$ & 0.087 & 0.103  & 0.028 & 0.214  & 0.325 & 0.211  & 0.009 & \\

\hline
$x_i$  & 0.814 & 1.485  & 0.451  & 1.475  & 0.445 & 1.796  & 0.807 &  \\
 $\hat \lambda_i$ & 0.187 & 0.249  & 0.308 & 0.249  & 0.312 & 0.356  & 0.187 & Step 2 \\
 $\hat w_i$ & 0.286 & 0.278  & 0.117 & 0.0882  & 0.0628 & 0.004  & 0.141 & \\

\hline
$x_i$   & 0.814 & 1.485  & 0.451 & --  & -- & --  & -- &  \\
 $\hat \lambda_i$ & 0.187 & 0.249  & 0.308 & --  & -- & --  & -- & Step 3 \\
 $\hat w_i$ & 0.426 & 0.369  & 0.183 & --  & -- & --  & -- & \\

\end{tabular}
\end{table}

The whole numerical experiment and unfolding was repeated ten times.
The number of obtained final components varied between
three and six. The obtained $p$-values show no clear
deviation from an evenly distribution between zero
and unity, indicating that the measured distributions
are typically reasonably well described by the folded estimates
of the true distribution -- supporting the validity of
the unfolding approach.

\subsection{Strongly varying one-sided distribution}
\label{sec:example2}
\begin{figure}[H]
\centering
\includegraphics[width=0.8\textwidth]{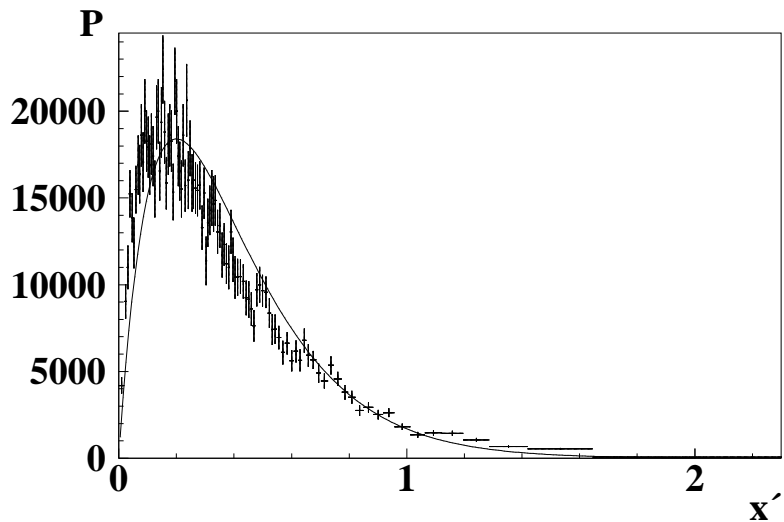}
\caption{The  histogram of the measured
         distribution $\bm{P}$\/ based on a sample of 10\,000 events generated
         for the true distribution. The true distribution $p(x)$\/
         is shown by the curve.}
\label{fig:resaccept2}
\end{figure}

In this example the above method is applied to unfold a strongly
varying one-sided PDF. The true distribution, defined in the
range $[0, +\infty)$, is
\begin{equation}
p(x) \propto xe^{-5x} \;.
\label{testform2}
\end{equation}
Let us represent the true value $x$\/ as a function of two variables $u$\/ and $v$,
$x=\sqrt{u^2+v^2}$, with $u=x \cos(\phi)$\/ and $v=x \sin(\phi)$, with
the angular variable $\phi$\/  uniformly distributed in $[0, 2\pi)$.
The reconstructed value $x'=\sqrt{u'^2+v'^2}$\/ is obtained from $u'$\/
and $v'$, defined as independent random variables with normal distributions
$\mathcal{N}(u,(0.5u)^2)$\/ and  $\mathcal{N}(v,(0.5v)^2)$ respectively.
Here we do not present an analytical formula for the resolution function
$R(x'|x)$, but notice that it is a generalisation of the Rice distribution.
An example for the measured distribution obtained by simulating a sample
of $N=10\,000$\/ events is presented in Fig.~\ref{fig:resaccept2}.
\vspace *{-0.9 cm}
In general the choice of PDFMs has should be adapted
to the problem at hand, i.e. symmetric gaussian PDFMs as used in the
previous example are not directly suitable for this kind of unfolding
problem. The GMM fit model can, however, be used after transforming the
problem such that the true distribution becomes approximately gaussian in
shape. The Box-Cox transformation \cite{sakia}
 \begin{equation}
   x^{\lceil\mu\rfloor} = \left\{ \begin{array}{l}
                   (x^\mu-1)/\mu \quad \mbox{for $\mu \neq 0$} \\
                   \ln x  \qquad \qquad  \mbox{for $\mu=0$}
                   \end{array}\right.
\end{equation}
is appropriate for this case. After transforming the unfolding result back
to the original variables, the entire procedure is equivalent to using a
PDFMs of the form
 \begin{equation}
  K(x;x_i,\lambda,\mu)
   = \frac{1}{\lambda \sqrt{2 \pi}} \exp\left(
               {-\frac{(x^{\lceil\mu\rfloor}-x^{\lceil\mu\rfloor}_i)^2}{2\lambda^2}}
                    \right)x^{(\mu-1)} \;.
\label{pdfbc}
 \end{equation}
For the determination of the matrix $\bm{\mathrm{Q}}$\/ a sample of 1\,000\,000
Monte Carlo events was simulated. The true distribution was taken to be uniform
and the response of the PDFM was calculated by weighting the Monte Carlo events
with weights proportional to the value of respective PDFMs \cite{sobol}. In the
first step an initial set of 400 PDFMs was used with positions $x_i$ uniformly distributed over the interval $[0, 2]$.
The transformation parameter $\mu=0.25$\/ was used, which leads to a
transformed PDF $P(x'^{\lceil\mu\rfloor})$\/ with skewness close to 0.

\begin{figure}[H]
\centering
\vspace *{-1.5cm}\includegraphics[width=1.\textwidth]{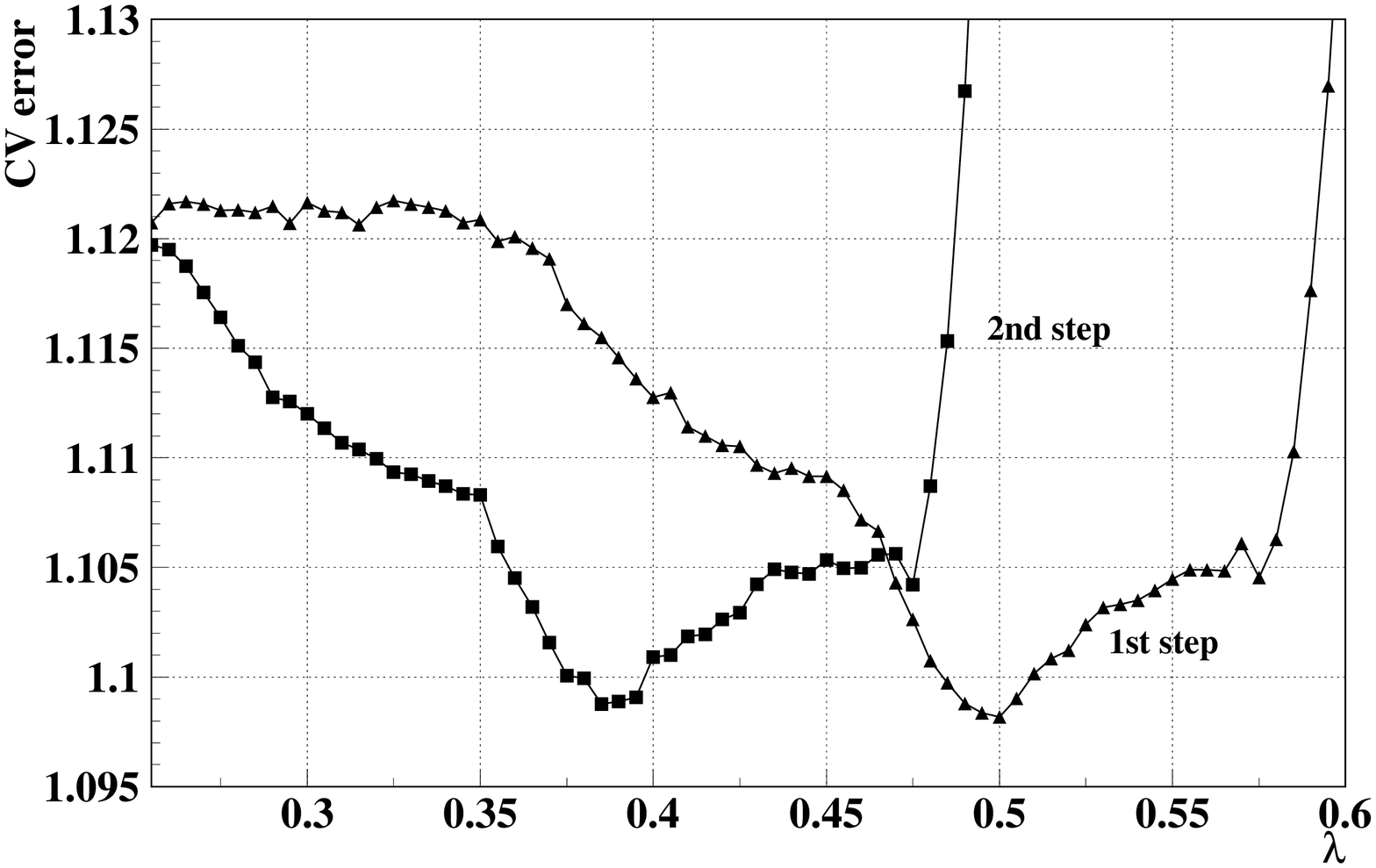}\\ \vspace *{-1.5 cm}
\includegraphics[width=1.\textwidth]{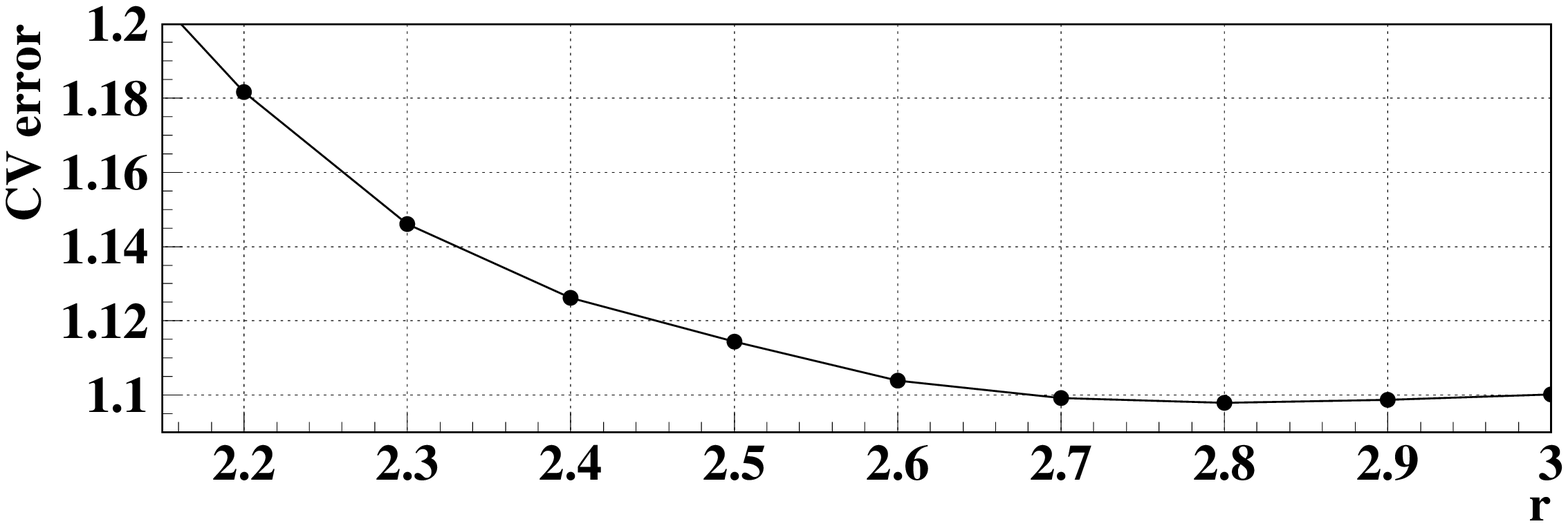}
\vspace *{-1. cm}
\caption{Cross-Validation errors  for different values of $\lambda$ (top) and
Cross-Validation errors for different values of $r$ ($\lambda=0.39$) (bottom).}
\label{fig:cross2}
\end{figure}

\begin{figure}[H]
\begin{center}$
\begin{array}{cc}
\includegraphics[width=9cm]{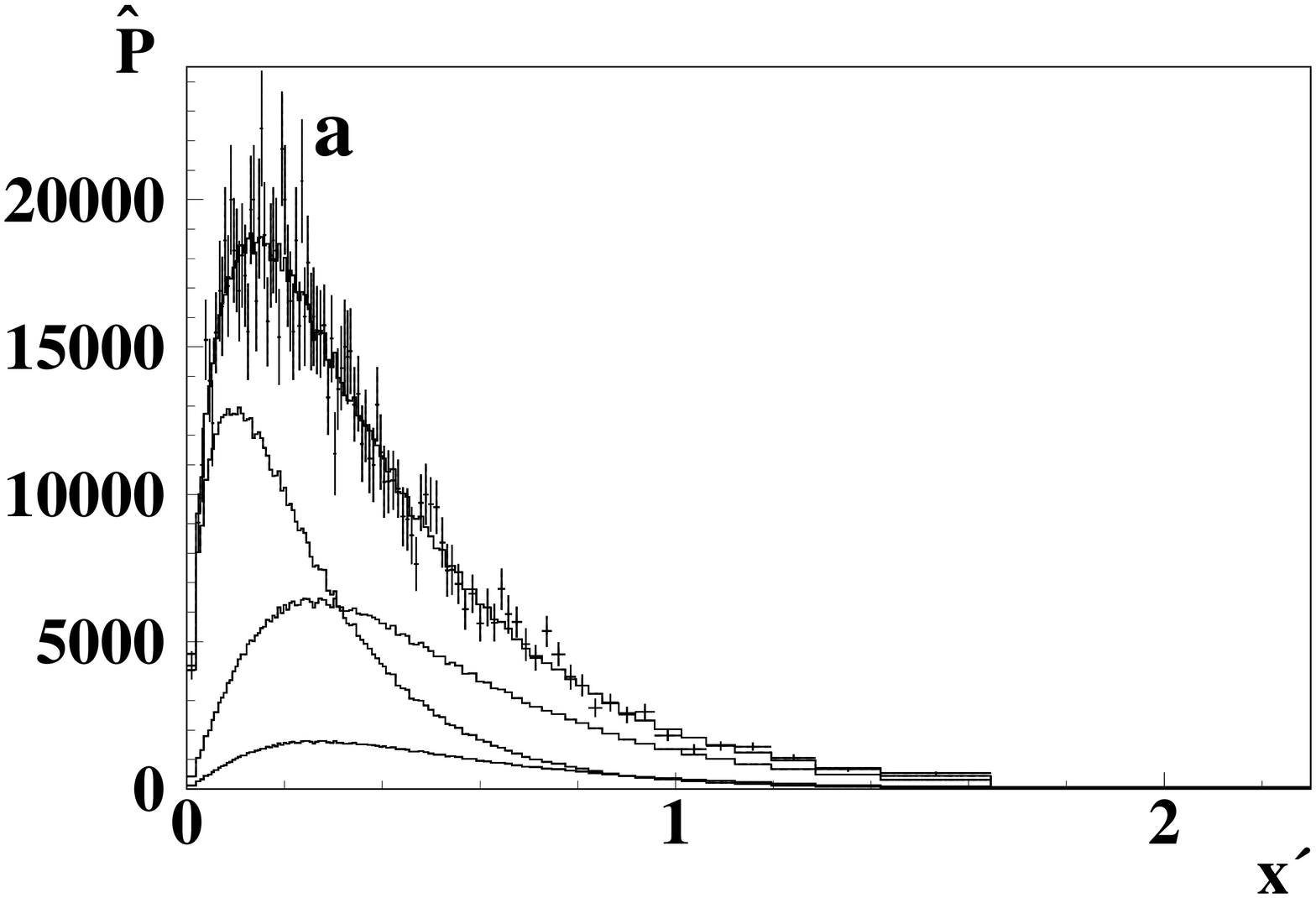} &
\vspace *{-1.1 cm} \hspace *{-0.6cm}\includegraphics[width=5.3cm]{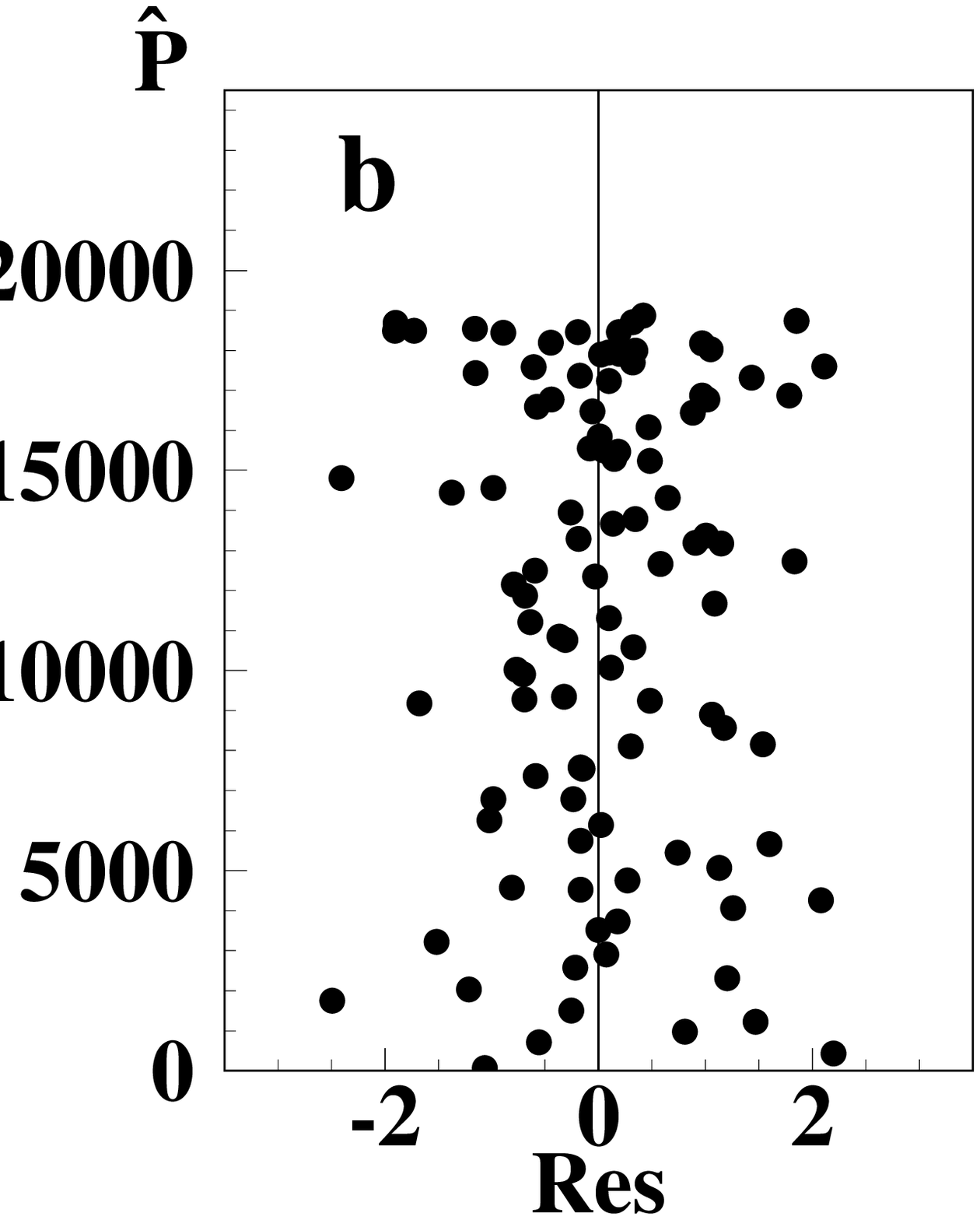} \vspace *{0.3cm}\\
\includegraphics [width=9cm]{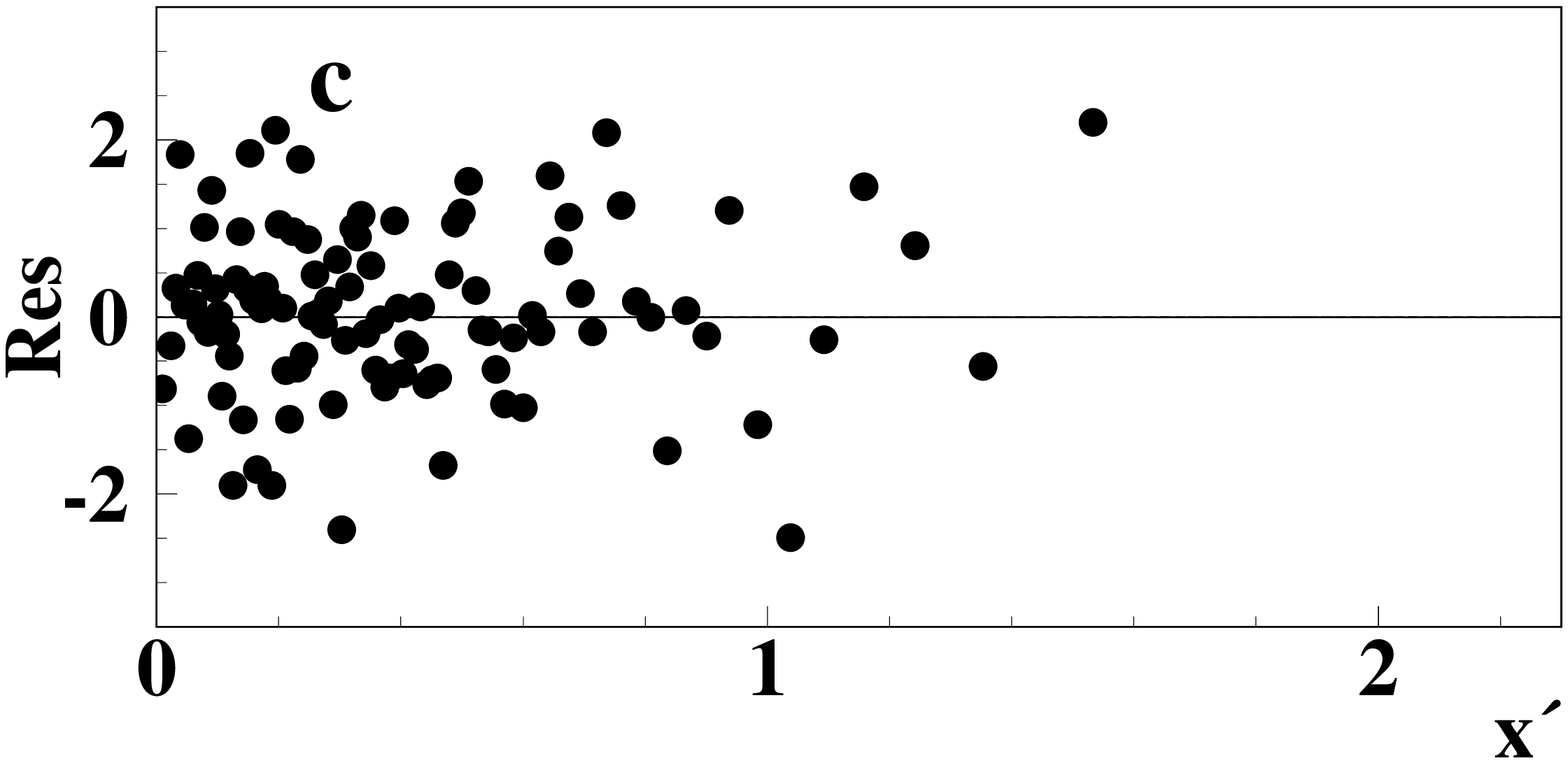}&
\vspace *{-1.1 cm} \hspace *{-0.6cm}\includegraphics[width=5.3cm]{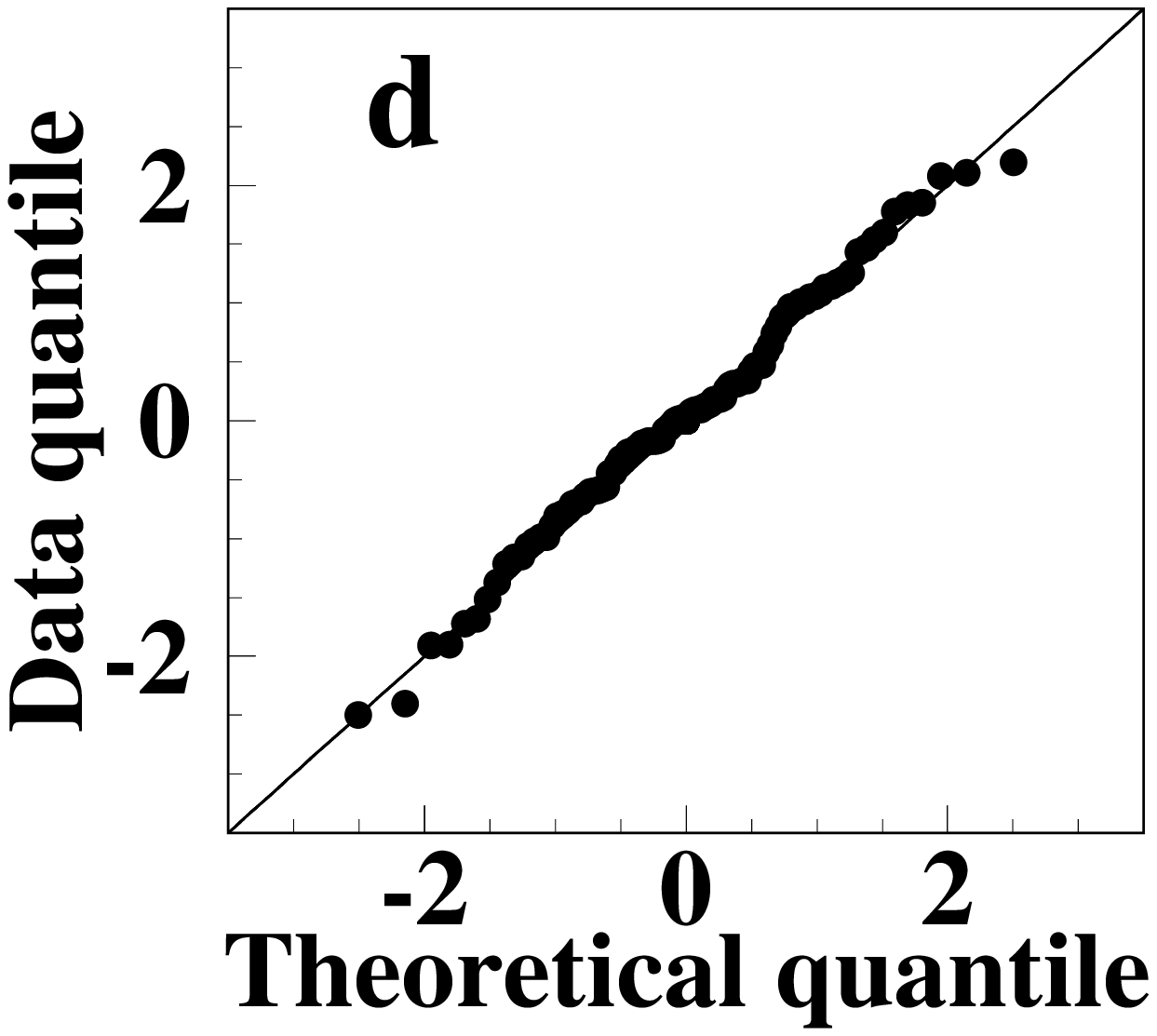}
\end{array}$
\end{center}
\vspace *{0.5cm}
\caption{Illustration of the quality of the unfolding result:
        (a) folded  estimate of the true distribution $\hat{\bm{P}}$ (solid histogram) compared to the measured distribution $\bm{P}$; (b) normalised residuals
        of the fit as a function of $\hat{\bm{P}}$; (c) normalised
        residuals as a function of $x'$; (d) quantile-quantile-plot
        for the normalised residuals.}
\label{fig:quality2}
\end{figure}

\begin{figure}[H]
\begin{center}$
\begin{array}{cc}
\vspace*{-1.92cm}\includegraphics[width=2.99in]{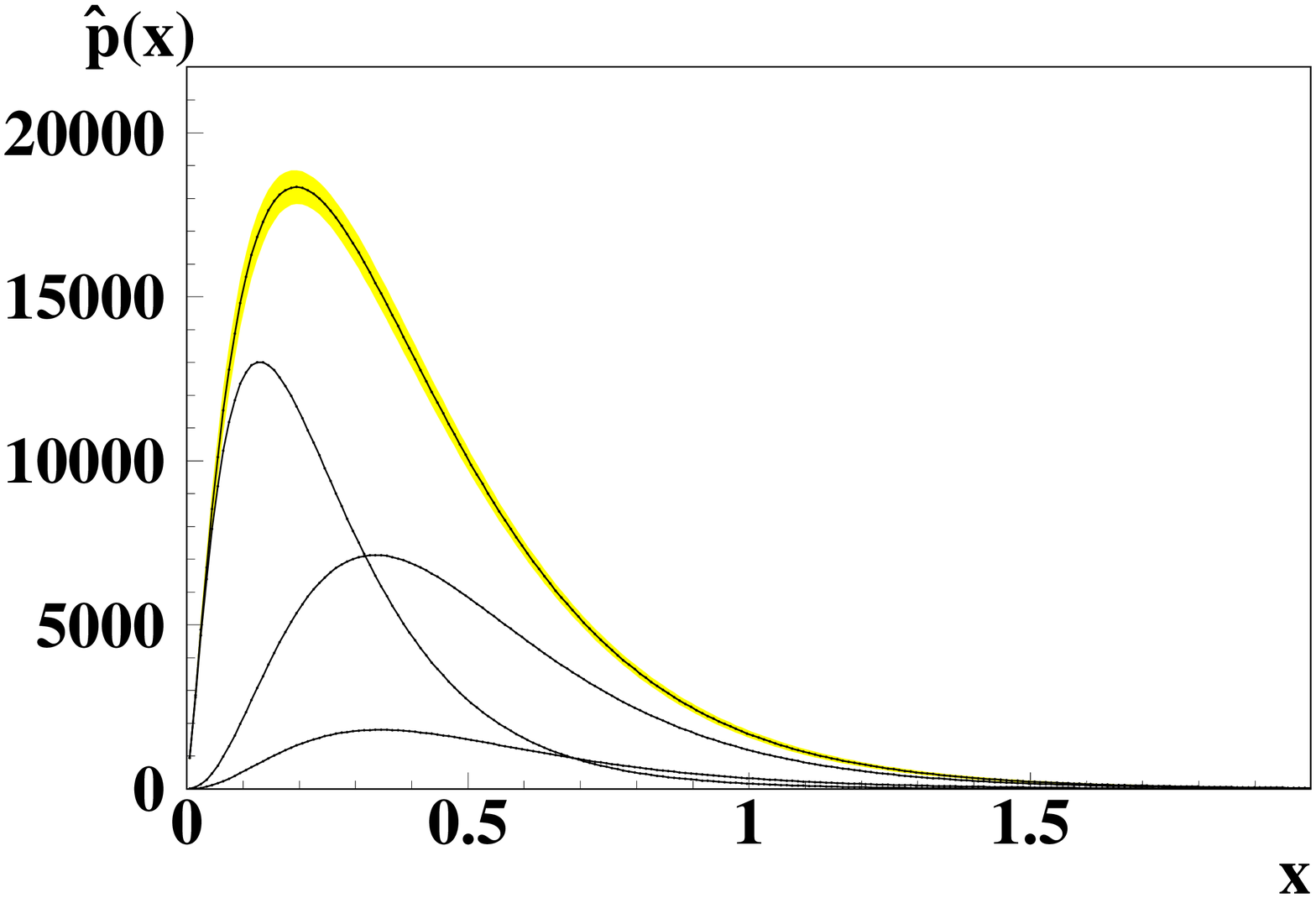} &
\hspace*{-1.82cm}\includegraphics[width=2.99in]{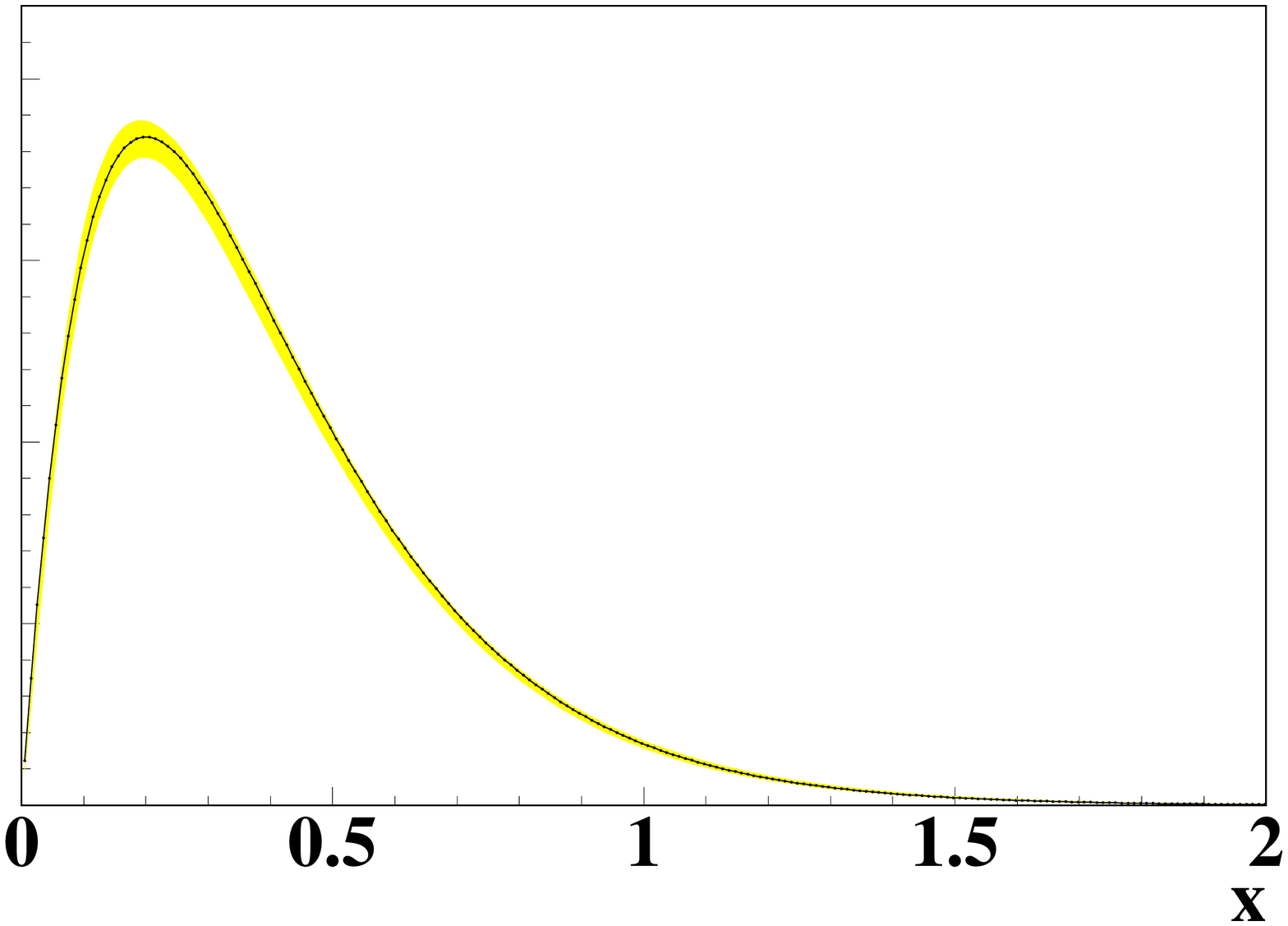}\\
\end{array}$
\end{center}
\vspace*{1.0cm}
\caption{Components of the unfolded distribution and the unfolded
         distribution $\hat{p}(x)$\/ given by the sum of the components
         with $\pm 2\delta(x)$\/ interval (left) and the two standard deviations band overlaid with the true distribution $p(x)$\/ (right).}
\label{fig:unfolded2}
\end{figure}

\begin{figure}[H]
\begin{center}
\includegraphics[width=4.5in]{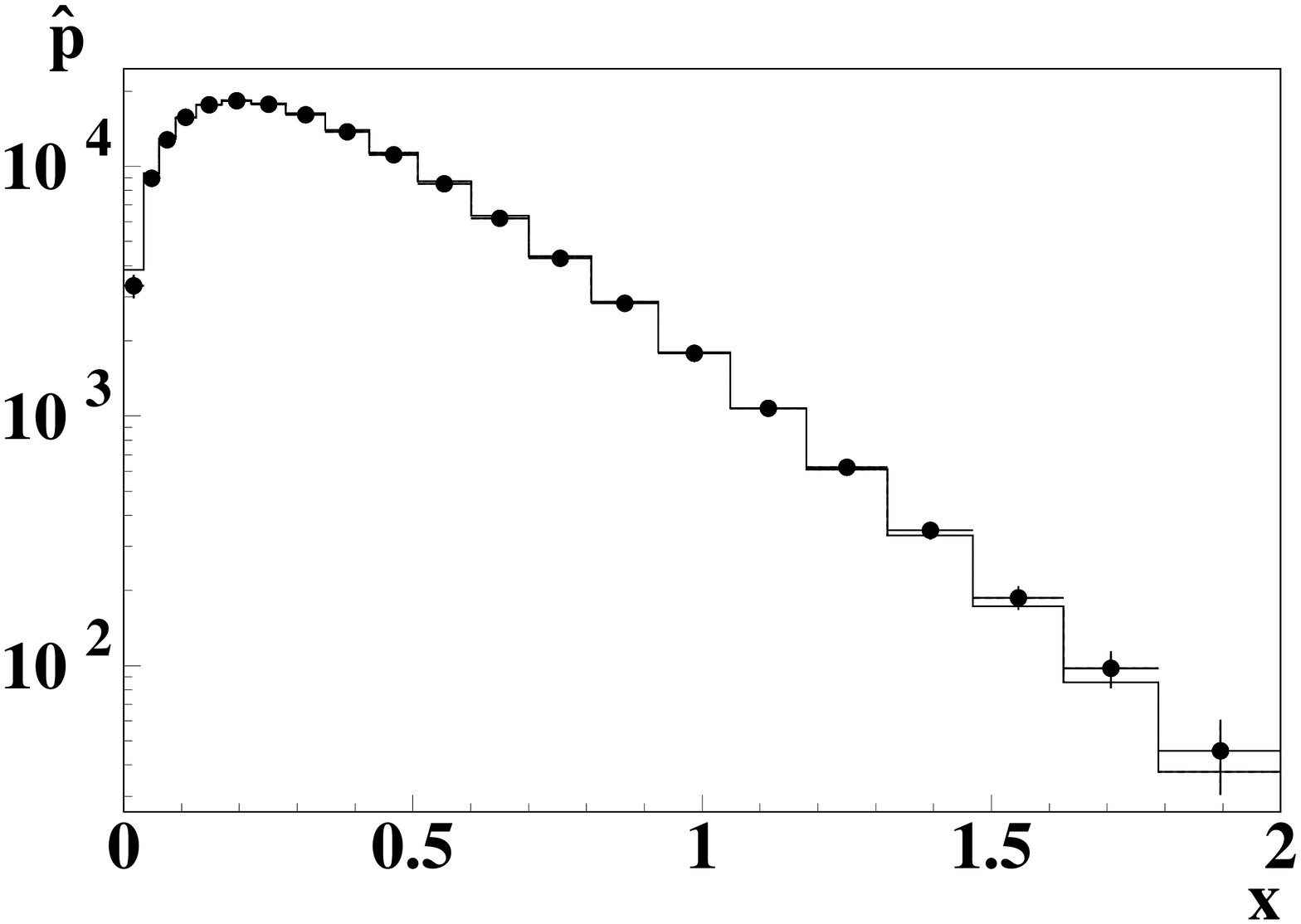}
\end{center}
\caption{Binned representation of the unfolding results $\hat{p}_i$\/
         for $m=21$\/ bins. The vertical error bars denote the standard
         deviations $\delta_{i}$. The histogram shows the true bin contents $p_i$.}
\label{fig:unfoldedhist2}
\end{figure}

As shown in Fig.\,\ref{fig:cross2}, the constant width parameter
$\hat\lambda_{i}=\hat \lambda=0.5$\/ provides the minimum value for the
Cross-Validation error $CV(\lambda)$. In the second step $\hat\lambda_{i}$\/
is not constant  but inversely proportional to the square root of the
unfolded density obtained in the first step in order to have better
smoothing in regions of low statistics. Here one finds a preferred value of
$\hat\lambda=0.39$. In this case the Cross-Validation error does not
improve. In the third step finally a best value for the garrote parameter
$r=2.8$\/ is found for $\hat \lambda=0.39$. The minimun, however, is not
very pronounced. These two parameters are used for the final calculation
of the unfolded distribution. Only three terms are retained for the estimate
of the true distribution.
Figure \ref{fig:quality2} illustrates the quality of the fit. No structure
in either of the control plots is observed. The $p$-value from the test
for the comparison of the histogram of the measured distribution $\bm{P}$\/
and the fitting histogram $\hat{\bm{P}}$, Fig.~\ref{fig:quality2}(a),
is $p=0.43$. The components of the unfolding results are shown together
with the estimate $\hat{p}(x)$\/ in Fig.~\ref{fig:unfolded2}. Also shown
are the standard deviation bands $\pm 2\delta(x)$\/ compared to the true
distribution $p(x)$. The binned presentation of the unfolded distribution
shown in Fig.\,\ref{fig:unfoldedhist2} was done with $n=21$\/ bins.
\vspace{-1.cm}
\newpage
\vspace *{-1.cm}

\subsection{Two-dimensional unfolding}
\label{sec:example4}
The method presented in this paper is also directly applicable to
multidimensional cases. Here a two-dimensional example is given. The
true distribution defined on the two-dimensional domain
$x,y \in [10,40]\times[10,40]$\/ and represented  by a sum of three
bivariate gaussian PDFs $p_g(x, y; \mu_x, \mu_y, \sigma_x,\sigma_y, \rho)$,
with $\mu_x, \mu_y$\/ the expectation values, $\sigma_x, \sigma_y$\/ the
standard deviations and $\rho$\/ the correlations coefficient. The
actual density is given by
\begin{equation}
\begin{split}
p(x,y) &  \propto 4\,   p_g(x, y; 20.5, 25.5,  ~4.0, ~4.0, 0.5) \\
       &    +     ~\,\, p_g(x, y; 30.5, 25.5,  ~3.0, ~3.0, 0.0) \\
       &    +     5\,   p_g(x, y; 35.0, 23.0,  20.0, 30.0, 0.0) \;.
\end{split}
\label{testform4}
\end{equation}
The experimentally measured distribution is obtained by
\begin{equation}
 P(x',y') \propto \int_{10}^{40} \int_{10}^{40} p(x,y)R(x',y'|x,y)dxdy \;,
\end{equation}
with a resolution function
\begin{equation}
\begin{split}
  R(x',y'|x,y) & \propto \quad\;\;\,\, p_g(x', y'; x, y, 1.0, 1.0, 0.0) \\
               &    +    ~\,0.5     \, p_g(x', y'; x, y, 2.5, 2.5, 0.0) \\
               &    +       0.05    \, p_g(x', y'; x, y, 5.0, 5.0, 0.0) \;.
\end{split}
\label{res4}
\end{equation}

\begin{figure}[H]
\includegraphics[width=0.5\textwidth]{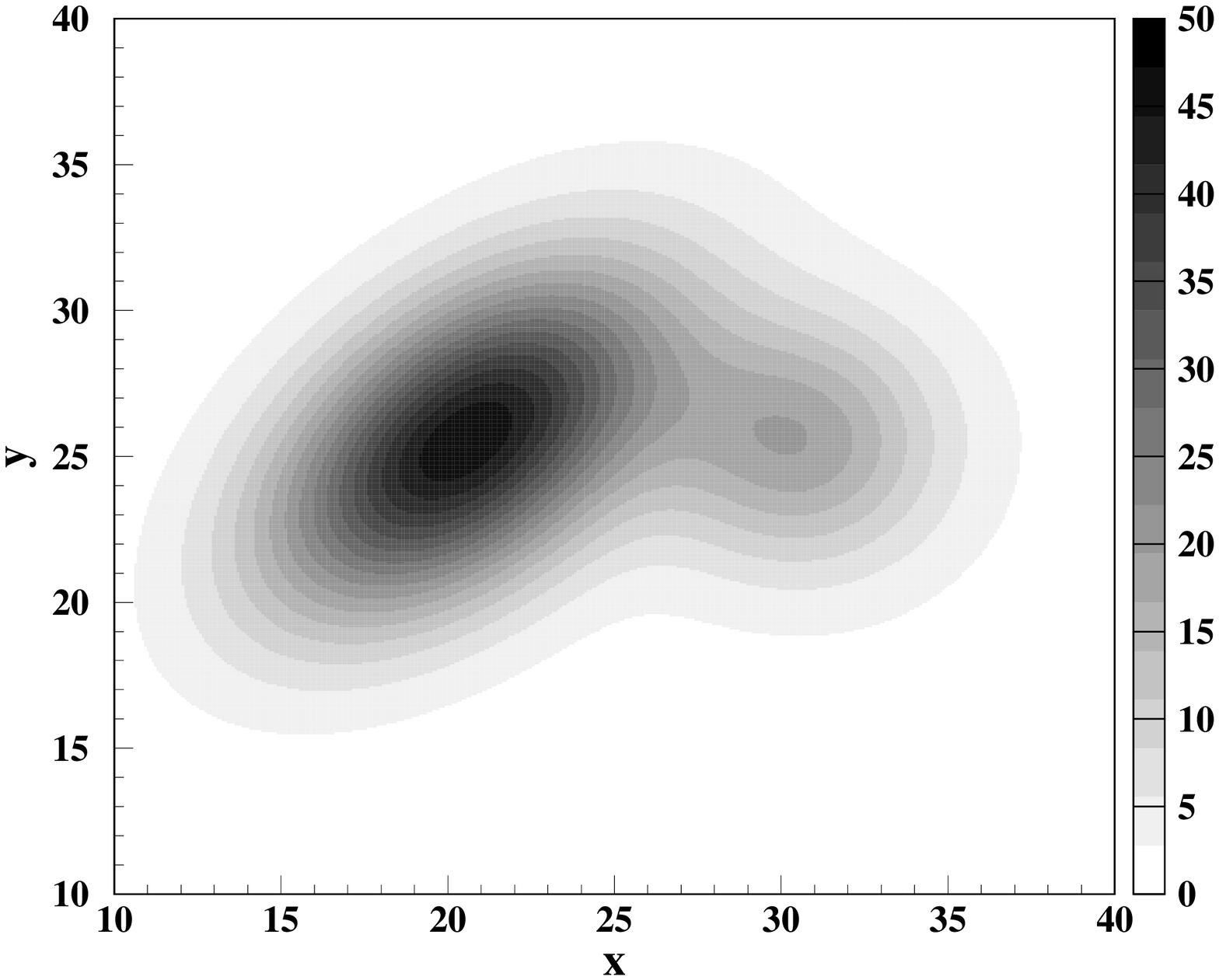}
\includegraphics[width=0.5\textwidth]{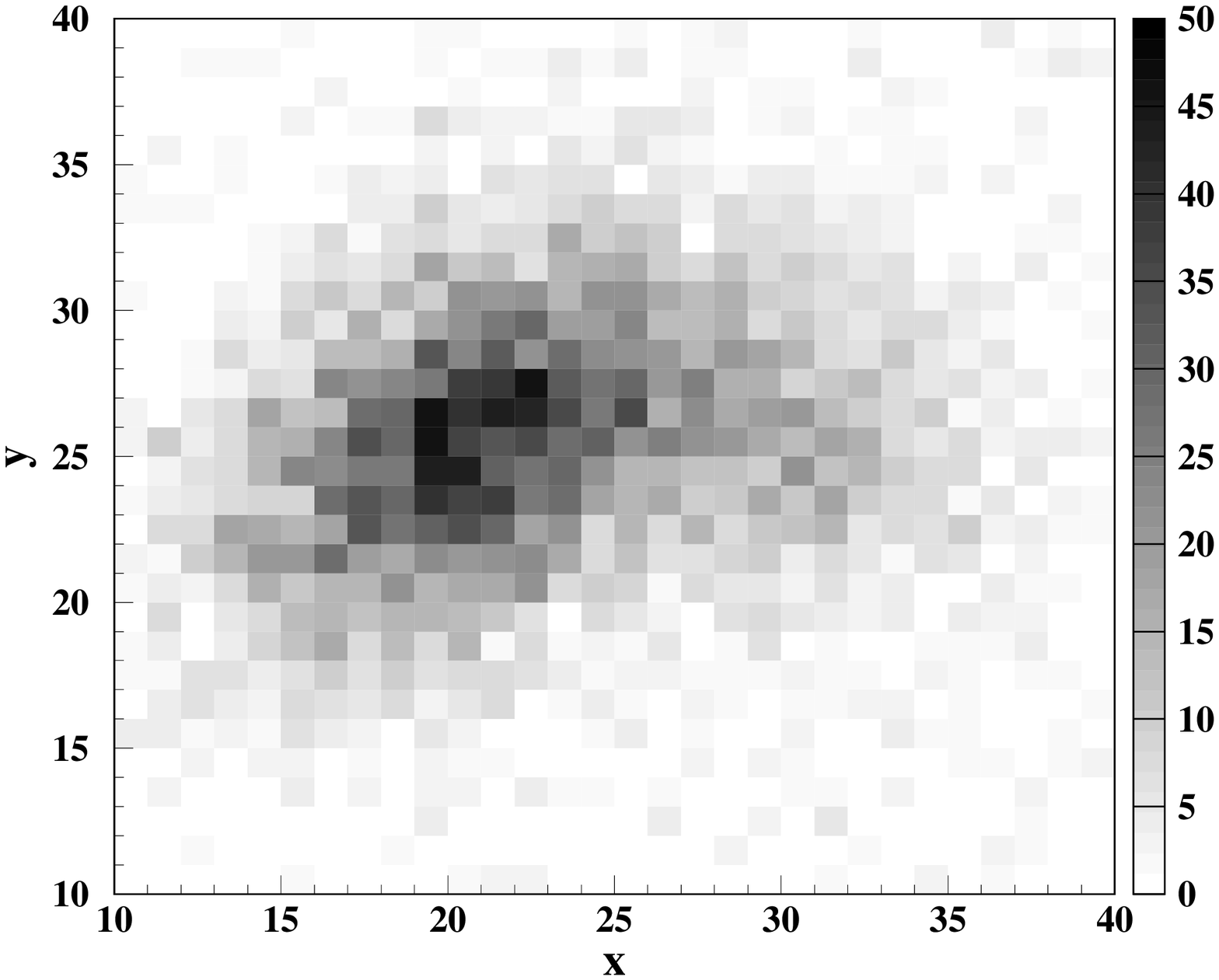}
\caption{The true distribution $p(x, y)$ (left) ) and histogram of the measured distribution $\bm{P}$\/ based on a sample of 10\,000 events (right).}
\label{fig:measured4}
\end{figure}

An example of a measured distribution is obtained by simulating
a sample of $N=10\,000$ events. In Figure\,\ref{fig:measured4} the true
density and the measured distribution are shown. While the dominant
component (first gaussian) is still clearly visible in the observed distribution, the weak component (second gaussian)  is barely discernible.

\begin{figure}[H]
\vspace *{-1.7cm}
\centering
\includegraphics[width=0.9\textwidth]{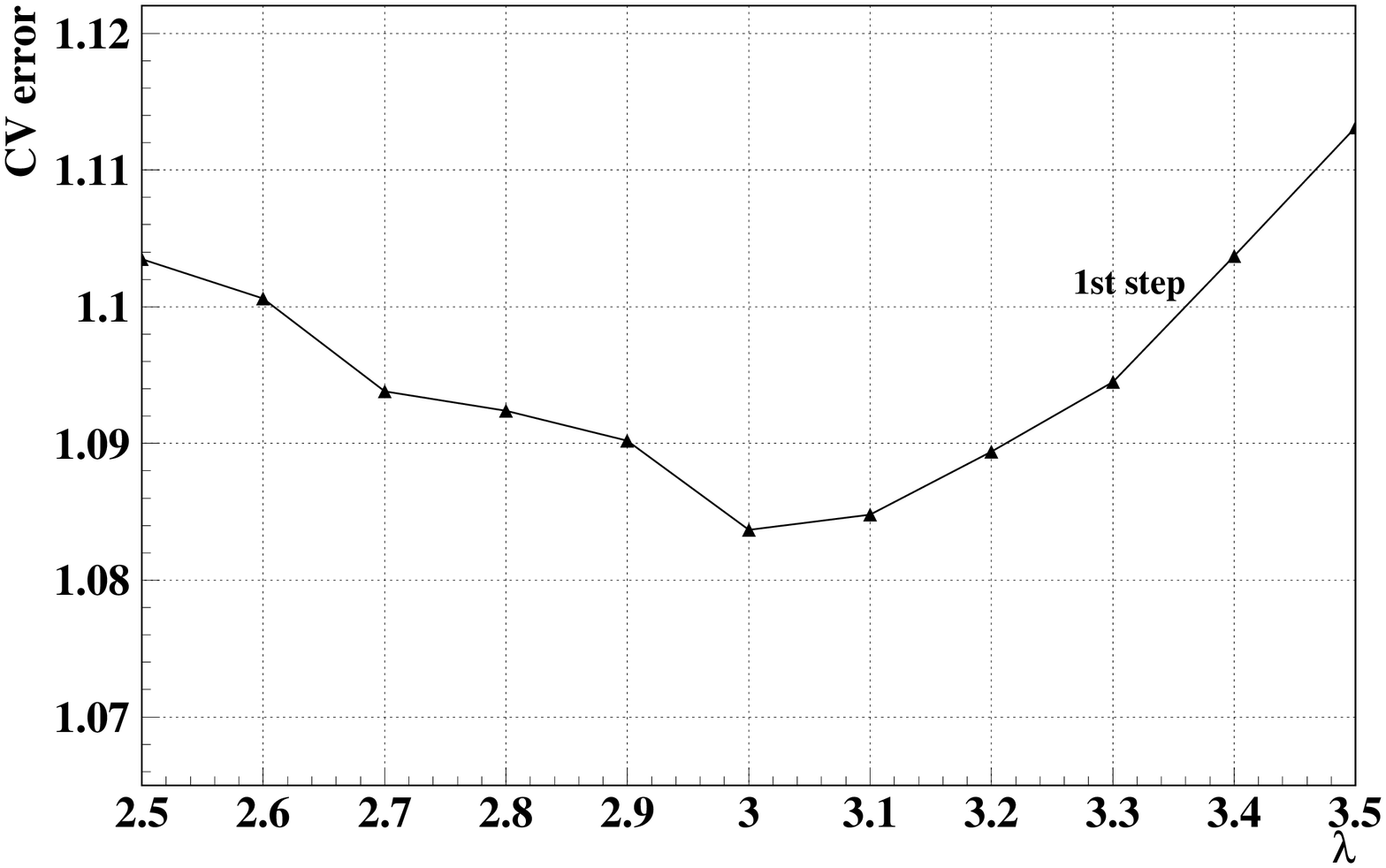}\\ \vspace *{-1.6cm}
\includegraphics[width=0.9\textwidth]{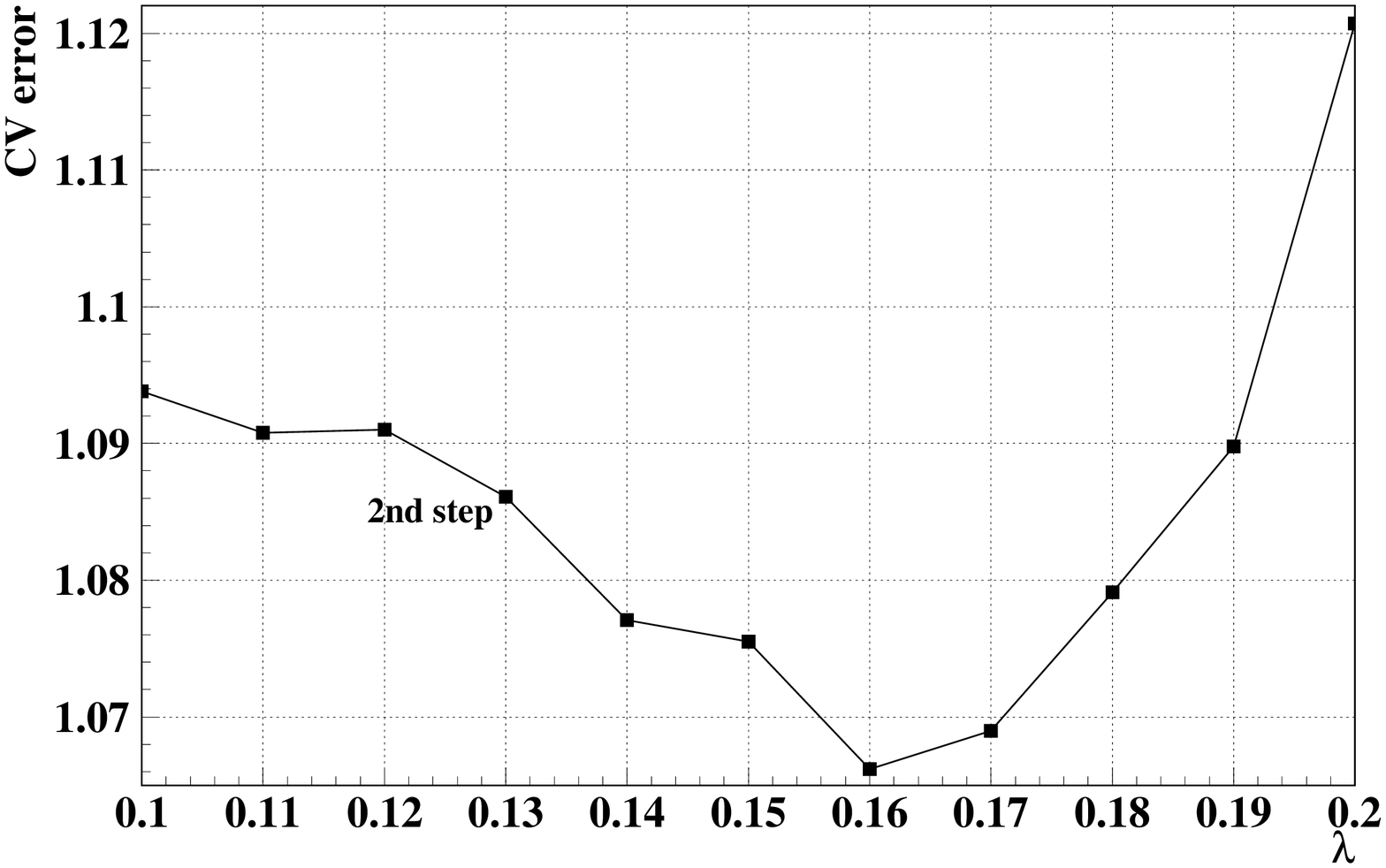}\\ \vspace *{-1.6cm}
\includegraphics[width=0.9\textwidth]{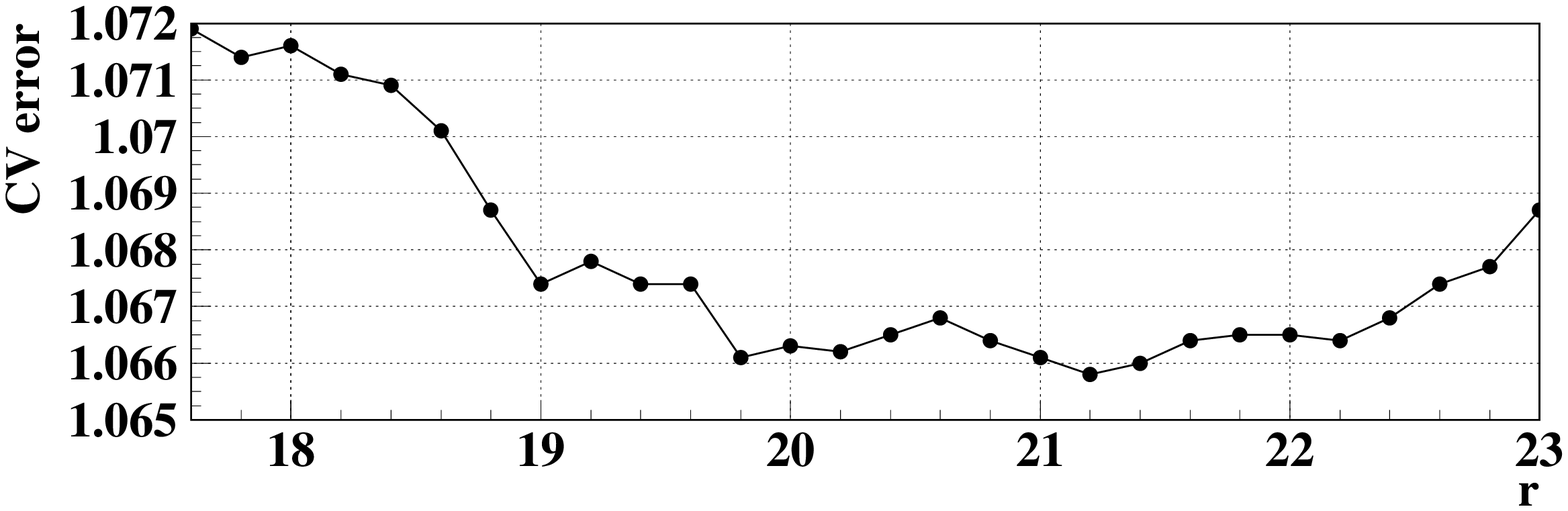}
\vspace *{-1.1cm}
\caption{Cross-Validation as a function of $\lambda$\/ in step 1 (top) and
         step 2 (middle), and as a function of the garrote parameter $r$\/
         for $\lambda=0.16$\/ from the second step (bottom).}
\label{fig:cross4}
\end{figure}

To use the existing software, a one-dimensional histogram was created
from the two-dimensional distribution by copying first the top row
from left to right then the 2nd row from  right to left and so on.
Adjacent bins of the one-dimensional histogram  with a low number
of events were merged to have at least 25 events per bin. The vector
of the histogram contents $\bm P$\/ finally used in the unfolding
procedure has $n=196$\/ components.
For the determination of the matrix $\bm{\mathrm{Q}}$\/ a sample of
1\,000\,000 Monte Carlo events was simulated. PDFMs were defined as circular symmetric gaussian probability density functions with three parameters, the expectation values $x_i, y_i$\/ and the standard deviation $\lambda_i$.
In the first step, a set of 400 PDFMs was used with positions $x_{i}, y_{i}$\/
uniformly distributed over the domain $x,y \in [10,40]\times[10,40]$. As shown
in Fig.\,\ref{fig:cross4}, using 5-fold Cross-Validation, an optimal value
$\hat\lambda_{i}=\hat \lambda=3.0$\/ is found. For the second step, with an
adaptive width $\hat\lambda_{i}$\/ inversely proportional to the square
root of the unfolded density obtained in the first step, the value
$\hat\lambda=0.16$\/ minimises the Cross-Validation error $CV(\lambda)$.
In the third step the optimal garrote parameter for $\hat\lambda=0.16$\/
is found to be $r=21.2$. These two parameters are used for the final
calculation of the unfolded distribution.

\vspace *{-0.5cm}
\begin{figure}[H]
\begin{center}$
\begin{array}{cc}
\includegraphics[width=9cm]{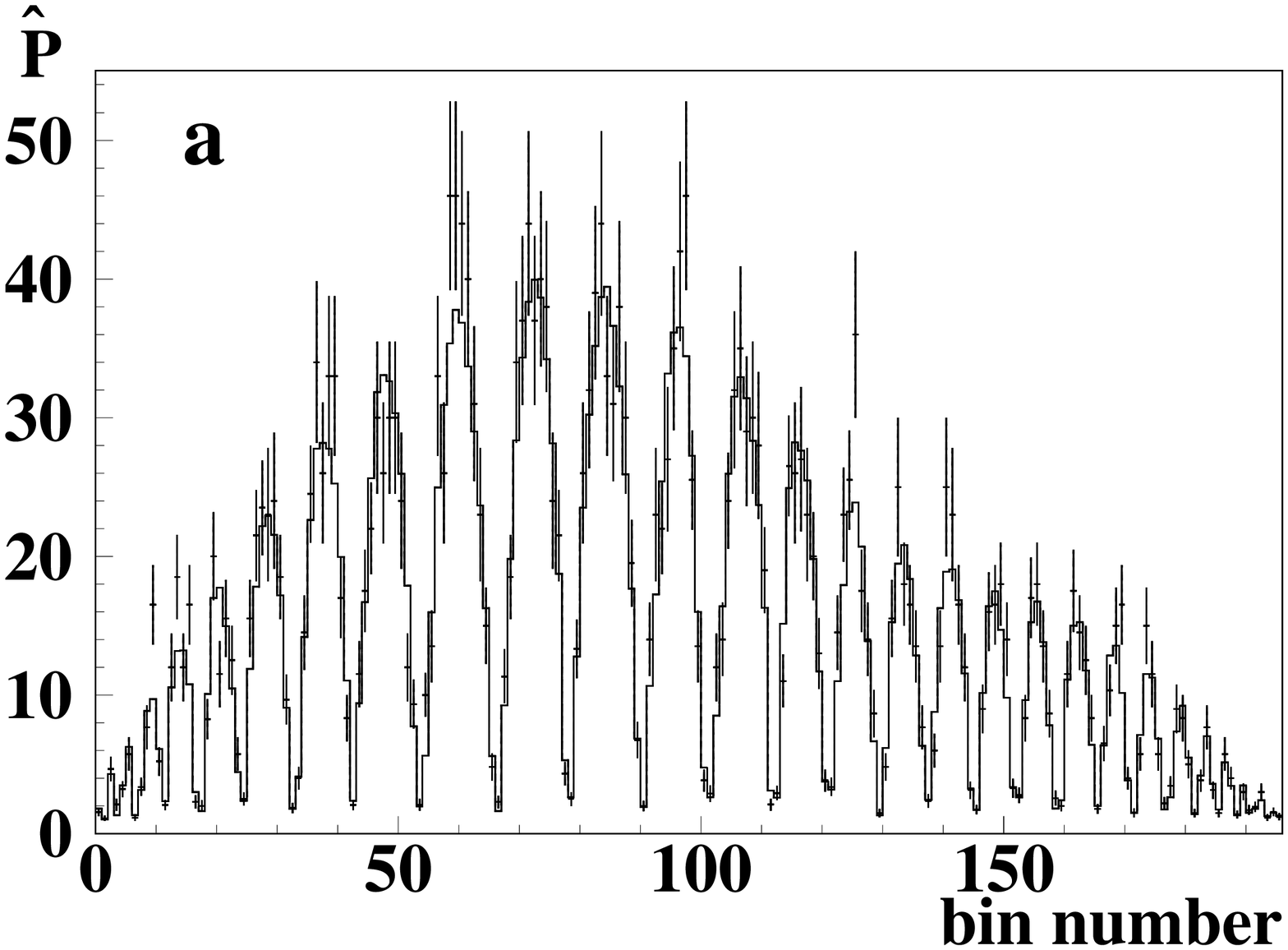} &
\vspace *{-1.1 cm} \hspace *{-0.6cm}\includegraphics[width=5.3cm]{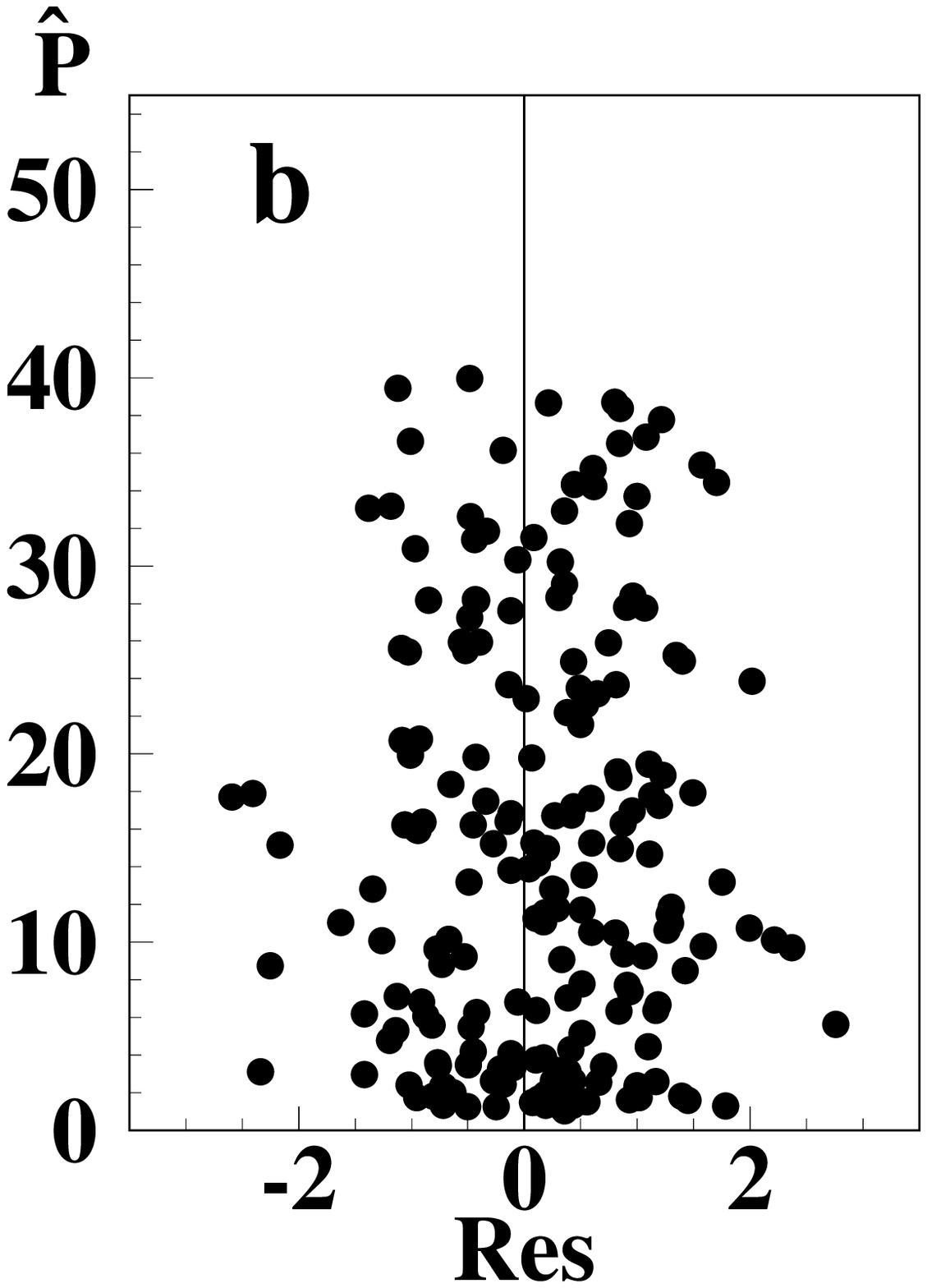} \vspace *{0.3cm}\\
\includegraphics [width=9cm]{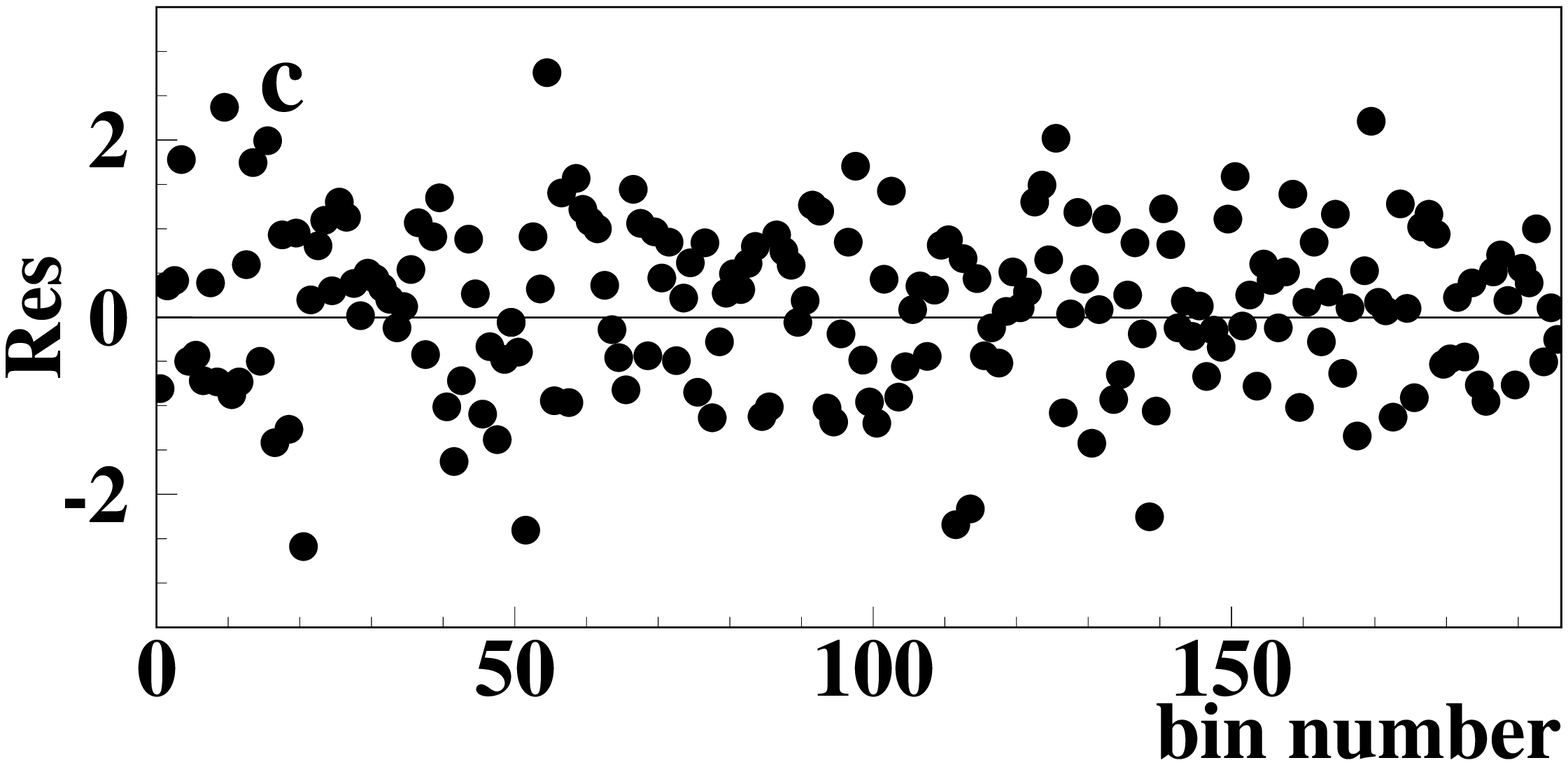}&
\vspace *{-1.1 cm} \hspace *{-0.6cm}\includegraphics[width=5.3cm]{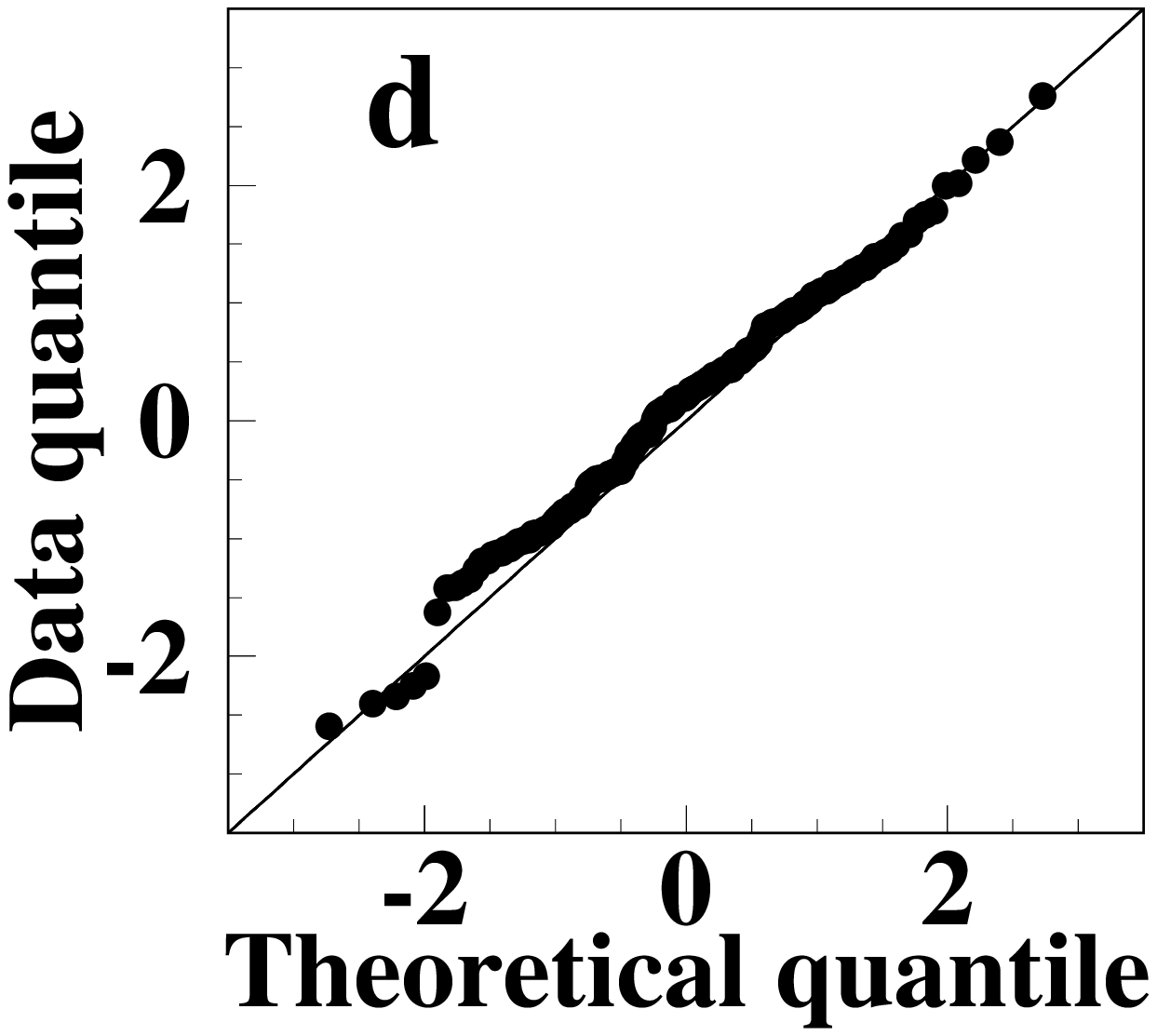}
\end{array}$
\end{center}
\vspace *{0.5cm}
\caption{Illustration of the quality of the unfolding result.
        (a) folded  estimate of the true distribution $\hat{\bm{P}}$ (solid histogram) compared to the measured distribution $\bm{P}$; (b) normalised residuals
        of the fit as a function of $\hat{\bm{P}}$; (c) normalised
        residuals as a function of bin number; (d) quantile-quantile plot
        for the normalised residuals.}
\label{fig:quality4}
\end{figure}

The quality of the unfolding result is illustrated by Fig.\,\ref{fig:quality4}.
It shows the mixture of the folded gaussian PDFMs, which approximates the
measured distribution, together with the analysis of the residual and the
quantile-quantile plot. No structure in either of the control plots is
observed. The $p$-value  for the comparison of the histogram of the measured
distribution $\bm{P}$\/ and the fitting histogram $\hat{\bm{P}}$\/ is $p=0.19$.

\begin{figure}[t!]
\includegraphics[width=0.5\textwidth]{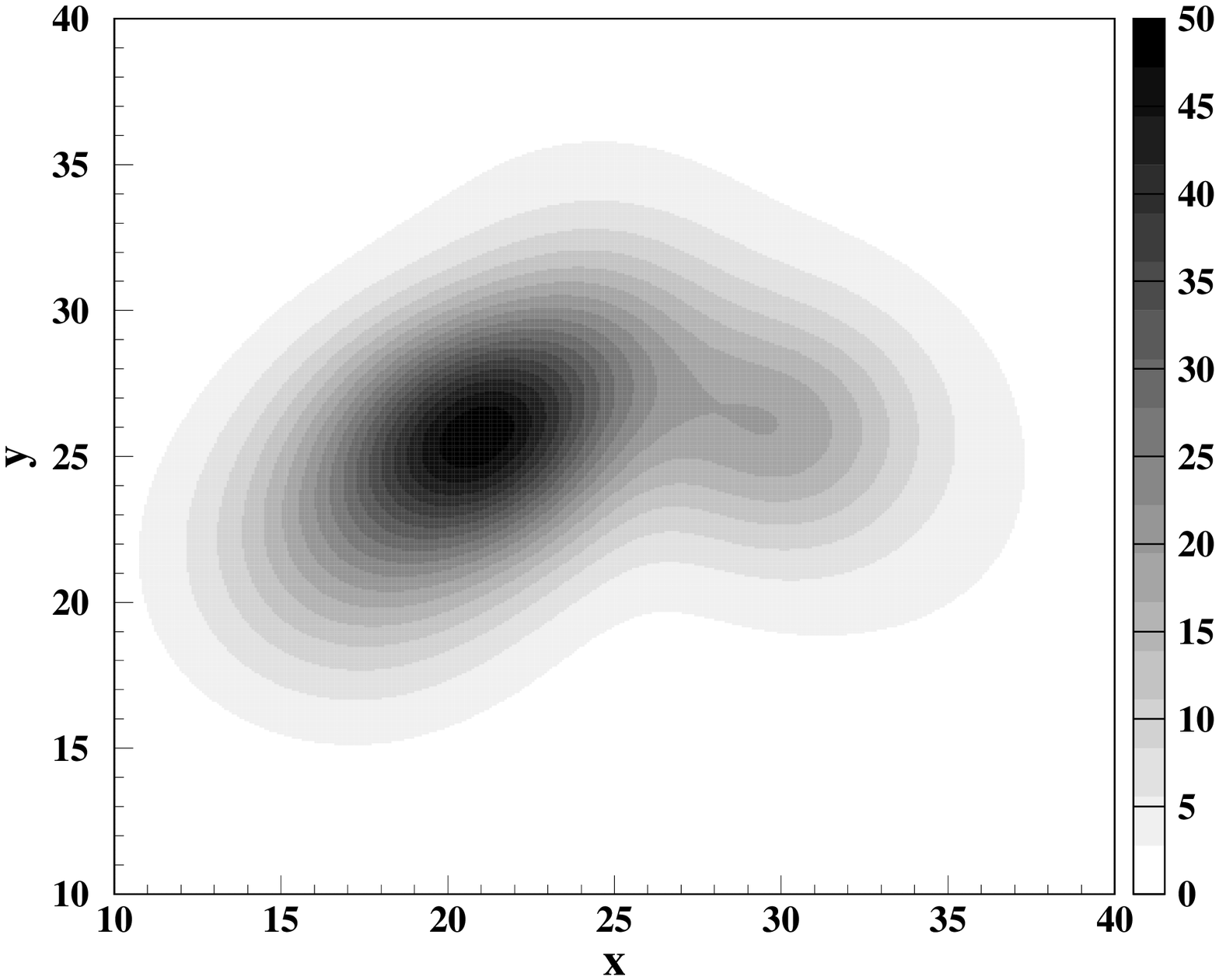}
\includegraphics[width=0.5\textwidth]{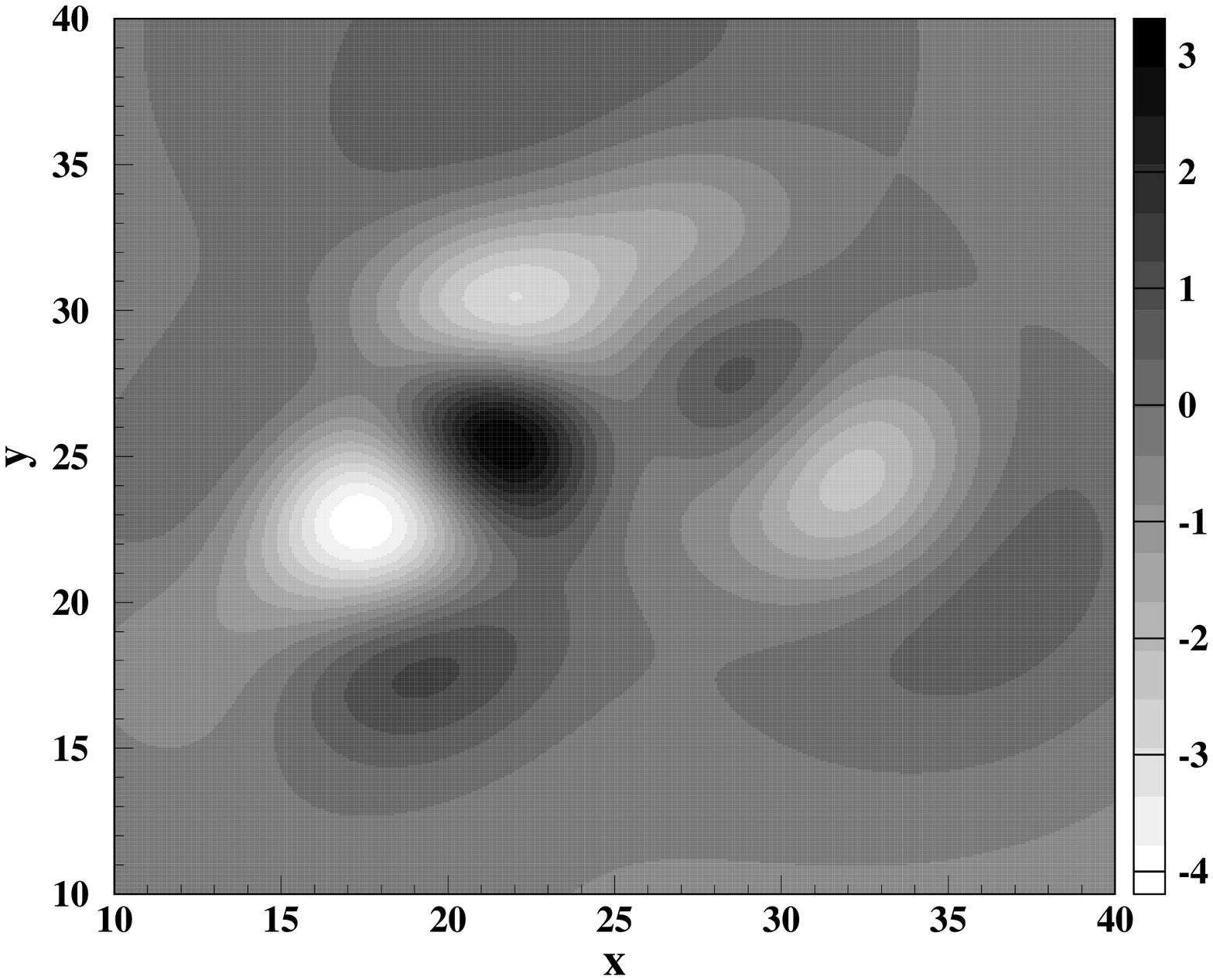}
\caption{The unfolded distribution $\hat p(x,y)$ (left) and
         the difference  between unfolded and true distribution
         $\hat{p}(x,y)-p(x, y)$ (right).}
\label{fig:unfolded4}
\end{figure}

The unfolded distribution $\hat p(x,y)$ is presented in
Fig.~\ref{fig:unfolded4} together with the difference between the
unfolded and the true distribution $\hat p(x,y)-p(x,y)$. One observes
that the true distribution, including its weak component, is well
reproduced by the unfolding result, with rather small deviations of
the unfolded distribution from the true one.

\section{Summary and conclusions}
\label{sec:conclusions}
A new method for unfolding the true distribution from data obtained
from detectors with finite resolution and limited acceptance is presented.
The method ensures smoothness and positivity of the result by
representing the true distribution as a weighted sum of smooth PDFs
(Mixture Densities Model). The standard deviation of the PDFs acts as
a regularisation parameter which determines the smoothness of the result.
The amount of smoothing is adjusted to the local statistical precision
of the data by scaling the width parameter inversely proportional to square root of the estimated density and the non-negative garrote method is used to eliminate insignificant terms in the solution. Cross-Validation is used
to determine optimal values of the regularisation and garrote
parameters. The method avoids discretisation of the true density entering
the integral equation, thereby avoiding quantisation errors for the true
distribution. The proposed procedure is directly applicable to
multidimensional unfolding problems. Numerical examples covering the
problems of unfolding a simple double-peak structure, a strongly varying
one-side distribution and a two-dimensional density were presented
to illustrate and to validate the procedure

\section*{Acknowledgements}
\noindent
The author would like to express very great appreciation
to Michael Schmelling for constant interest to this work and
many discussions stimulating the development of the method. The
author is also grateful to Markward Britsch for useful discussions
as well as for careful reading of the manuscript. Special thanks
are extended to the University of Akureyri and the MPI for Nuclear
Physics for support in carrying out the research.


\end{document}